# A memory-information window in finite-duration quantum control

*Doyeol (David) Ahn[1,2]**

[1]Department of Electrical and Computer Engineering,
University of Seoul, 163 Seoulsiripdae-ro, Tongdaimoon-gu, Seoul 02504, Republic of Korea
&
[2]Singularity Quantum Inc
895 Dove Street Second Floor, Newport Beach, CA 92660, USA

* Correspondence: dahn@uos.ac.kr

## Abstract

Environmental noise is a major source of decoherence and errors in quantum systems. In a non-Markovian environment, the noise has a finite correlation time, and the system can retain information about its previous interactions with the environment. Here, we investigate how such environmental memory can be observed through a finite-duration coherent quantum operation. Coherent control dynamically reshapes the stochastic coupling operator and thereby changes how two-time noise correlations affect the quantum process. For Ornstein-Uhlenbeck noise, we show analytically that two different limits reduce the information on the correlation time. In the short-memory limit, the leading dynamics depend on the integrated-noise combination $\sigma^2\tau_c$, where $\sigma^2$ is the noise variance and $\tau_c$ is the correlation time, making these parameters locally degenerate. In the long-memory limit, the stochastic field becomes quasi-static during the operation, and sensitivity to the correlation time is suppressed. Exact stochastic simulations for a driven single-qubit $X_\pi$ rotation and a two-qubit exchange gate show that independent correlation-time information becomes largest between these limits, when the environmental and control timescales are comparable. Noncommuting controls enhance this information, whereas commuting controls do not. Simulations at different gate durations $t_g$ further show that the information window approximately follows the dimensionless correlation ratio $\frac{\tau_c}{t_g}$. These results provide a timescale-matching principle for probing finite-correlation non-Markovian noise with coherent quantum control.

Noise and decoherence impose fundamental limitations on the coherent evolution of quantum systems. In quantum computing systems, these effects arise from interactions between qubits and their environment and determine the accuracy of quantum-state evolution and gate operations. A physical description of environmental noise is therefore important not only for estimating errors, but also for understanding how information about the environment is transferred to a controlled quantum system during an operation.1

Most descriptions of quantum noise characterize the environment through its strength or spectral distribution. In a non-Markovian process, the environment also retains memory over a finite correlation time, and the system can preserve signatures of earlier interactions with the stochastic field. Whether those signatures remain observable depends on how the correlation time compares with the characteristic timescale and temporal history of the coherent dynamics.2,3

Quantum noise spectroscopy, dynamical decoupling, filter-function methods and transfer-function approaches provide established ways to shape and interpret the temporal or spectral response of a controlled quantum system.4–9 In the present work, we consider a related but different question. Rather than asking only how strongly a control sequence responds to a component of the noise, we ask whether the noise strength and correlation time produce distinguishable changes in the complete quantum process generated during a finite-duration coherent operation.

This distinction separates environmental sensitivity from independent parameter identifiability. A quantum operation may be strongly affected by noise and may show a sizable derivative with respect to the correlation time, while that change remains nearly parallel to the process change produced by varying the noise strength. The operation then carries a clear noise signature, but the two environmental parameters cannot be inferred independently from the complete map. Establishing when the correlation-time direction becomes distinct from the strength direction is one of the central objectives of this study.

A finite-duration quantum operation provides only a restricted temporal window through which environmental dynamics can be sampled. If fluctuations decorrelate much faster than the gate duration, the operation mainly accumulates an integrated noise effect and fine correlation-time information is averaged out. If the fluctuations evolve much more slowly, the stochastic field appears approximately static during the operation: the gate can respond to its random amplitude while carrying little information about evolution outside the observation window. These different limits suggest an intermediate regime in which the correlation time may be most clearly resolved, without requiring its optimum to coincide exactly with the gate duration.

Earlier time-convolutionless (TCL) reduced-density-operator theory showed that coherent driving and stochastic-reservoir dynamics can interfere and thereby modify memory effects.[10,11] This formulation builds on the broader stochastic-Liouville description of fluctuating quantum dynamics.[12] In that formulation, coherent evolution between successive stochastic interaction events is not passive: it transforms the system operator through which the stochastic environment acts. The environmental correlation connects two events at different times, whereas the intervening coherent evolution determines the operator configurations associated with those events. In the present work, we revisit this mechanism by describing environmental fluctuations directly through a stochastic interaction Hamiltonian rather than introducing a harmonic-bath representation. This operator evolution provides the link between finite environmental memory and the duration and history of coherent control.

To quantify this effect, we examine the dependence of the complete process map on the noise strength and correlation time. The two parameters are independently identifiable only when their variations produce linearly independent changes in the quantum process. We therefore use the smallest singular value of the process Jacobian as an operational measure of the independent parameter direction. This construction makes the distinction explicit: absolute sensitivity to a change in correlation time does not by itself imply that correlation time can be separated from noise strength.

In this paper, we show that rapidly decorrelating Ornstein-Uhlenbeck (OU) noise produces a leading process that depends on $\sigma^2\tau_c$, making the noise strength and correlation time locally degenerate. For slowly varying noise, the stochastic field becomes quasi-static during the finite operation and loses independent sensitivity to $\tau_c$. Between these limits, independent correlation-time information becomes largest when the environmental and control timescales are comparable; the maximum is not assumed to occur at an exact equality of timescales.

We demonstrate this behavior using exact stochastic trajectory calculations for both a driven single-qubit $X_\pi$ rotation and a two-qubit isotropic exchange gate. We further use commuting controls as null tests, compare equal-endpoint pulse histories, and examine the scaling of the information window with the gate duration. The conclusions apply to the stationary finite-correlation Gaussian Ornstein-Uhlenbeck noise and the one- and two-qubit controlled systems studied here. These calculations separate the temporal connection imposed by the environmental correlation from the operator history generated by coherent evolution. We first establish this control-dependent two-time structure and then examine its consequences for the observability of environmental memory.

## Results

### Time-ordered coherent control reshapes stochastic interactions

We first formulate the mechanism for an arbitrary time-dependent coherent control, before introducing any model specialization. The total Hamiltonian is separated into coherent and stochastic contributions,

$$H(t) = H_s(t) + H_i(t), \quad H_i(t) = \sum_\alpha \xi_\alpha(t) A_\alpha \tag{1}$$

with zero-mean classical fluctuations and two-time correlations

$$\langle \xi_\alpha(t) \rangle = 0, \quad C_{\alpha\beta}(t,\tau) = \langle \xi_\alpha(t)\xi_\beta(\tau) \rangle \tag{2}$$

The exact coherent propagator is

$$u_s(t,\tau) = \mathcal{T}\exp[-i\int_\tau^t H_s(v)\mathrm{d}v] \tag{3}$$

and the Liouville superoperator and its propagator are defined by

$$\mathcal{L}_s(t)X = [H_s(t), X] \tag{4}$$

$$\mathcal{U}_s(t,\tau) = \mathcal{T}\exp[-i\int_\tau^t \mathcal{L}_s(v)\mathrm{d}v] \tag{5}$$

Their exact adjoint action is

$$\mathcal{U}_s(t,\tau)X = u_s(t,\tau) X {u_s}^\dagger(t,\tau) \tag{6}$$

No commutativity of the system Hamiltonian at different times is required for this identity; time ordering carries the complete control history. The proof is given in Supplementary Note I, Eqs. (S3)–(S6).

The stochastic operators in the control interaction picture are

$$\tilde{A}_\alpha(t) = {u_s}^\dagger(t,0)\, A_\alpha\, u_s(t,0), \quad \tilde{H}_i(t) = {u_s}^\dagger(t,0)\, H_i(t)\, u_s(t,0). \tag{7}$$

The environment supplies the temporal correlation, whereas coherent control determines which operator sector is sampled at each stochastic interaction time. This separation is the central physical mechanism considered here.

The second-order stochastic interaction structure (corresponding to the B-term in the original Supplementary Note II, Eqs. (S48)–(S50)) contains the schematic Liouville-space product

$$\mathcal{L}_i(t)\mathcal{U}_s(t,\tau)\mathcal{L}_i(\tau)\mathcal{U}_s^{-1}(t,\tau) \tag{8}$$

Thus two stochastic insertions are separated by coherent system propagation. Transforming to the control interaction picture converts the two insertions into the dressed operators at the two corresponding times. For a zero-mean classical stochastic interaction, the second-order TCL generator is

$$\mathcal{K}^{(2)}(t)X \;=\; -\int_0^t d\tau \,\langle[\tilde{H}_i(t),[\tilde{H}_i(\tau),X]]\rangle. \tag{9}$$

For a single stochastic channel with interaction Hamiltonian ξ(t)A, this becomes

$$\mathcal{K}^{(2)}(t)X \;=\; -\int_0^t d\tau \,C(t,\tau)\,[\tilde{A}(t),[\tilde{A}(\tau),X]]. \tag{10}$$

The full projection-operator algebra, including the P–Q decomposition, projected propagators, anti-time ordering, Born approximation and Eqs. (S18)–(S51), is retained in Supplementary Note II. The main text uses only the resulting two-insertion structure; the exact trajectory calculations below do not truncate the stochastic dynamics.

For conceptual bookkeeping only, we may write

$$K_O(t,\tau) \;=\; C(t,\tau)\,W_O(t,\tau;\,H_s,A) \tag{11}$$

Here $W_O(t,\tau;\,H_s,A)$ denotes the control- and observable-dependent weighting obtained by evaluating the dressed two-time operator structure for the chosen process observable O. It contains the dependence on the coherent propagator through $\tilde{A}(t)$ and $\tilde{A}(\tau)$, whereas C(t,τ) contains the environmental temporal correlation. This decomposition is introduced only for conceptual bookkeeping and is not intended to define $W_O$ as a standard literature object. This mechanism is summarized schematically in Fig. 1.

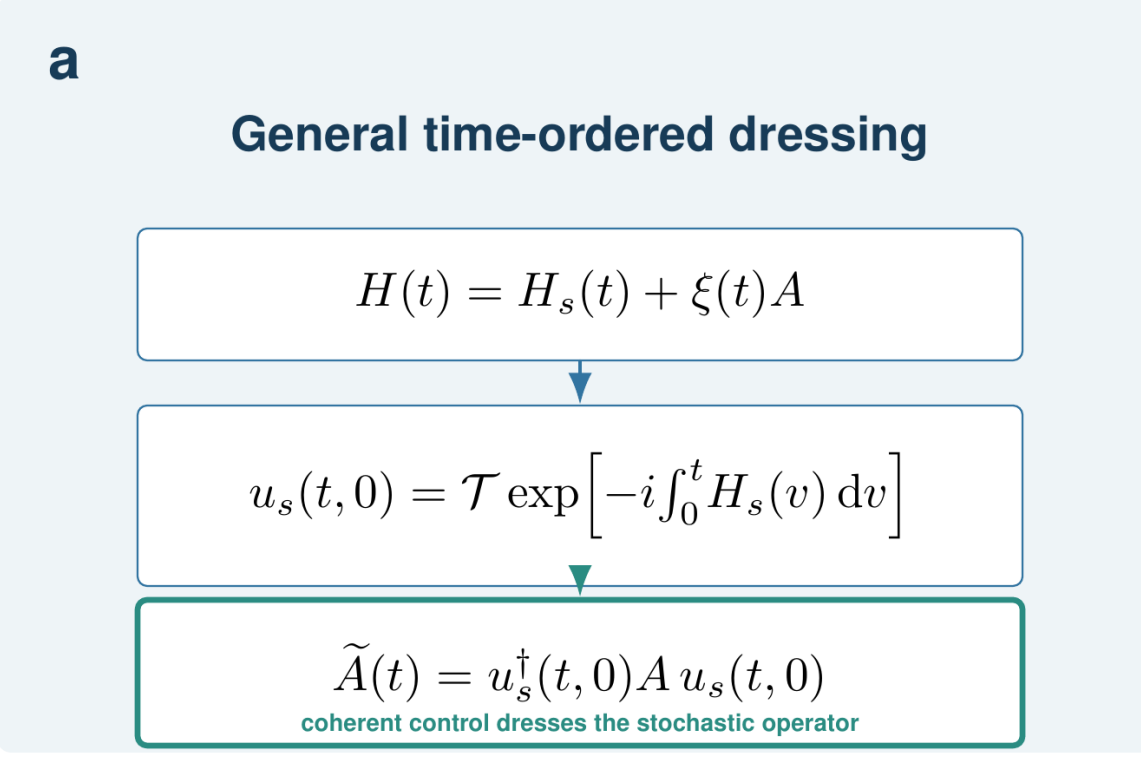

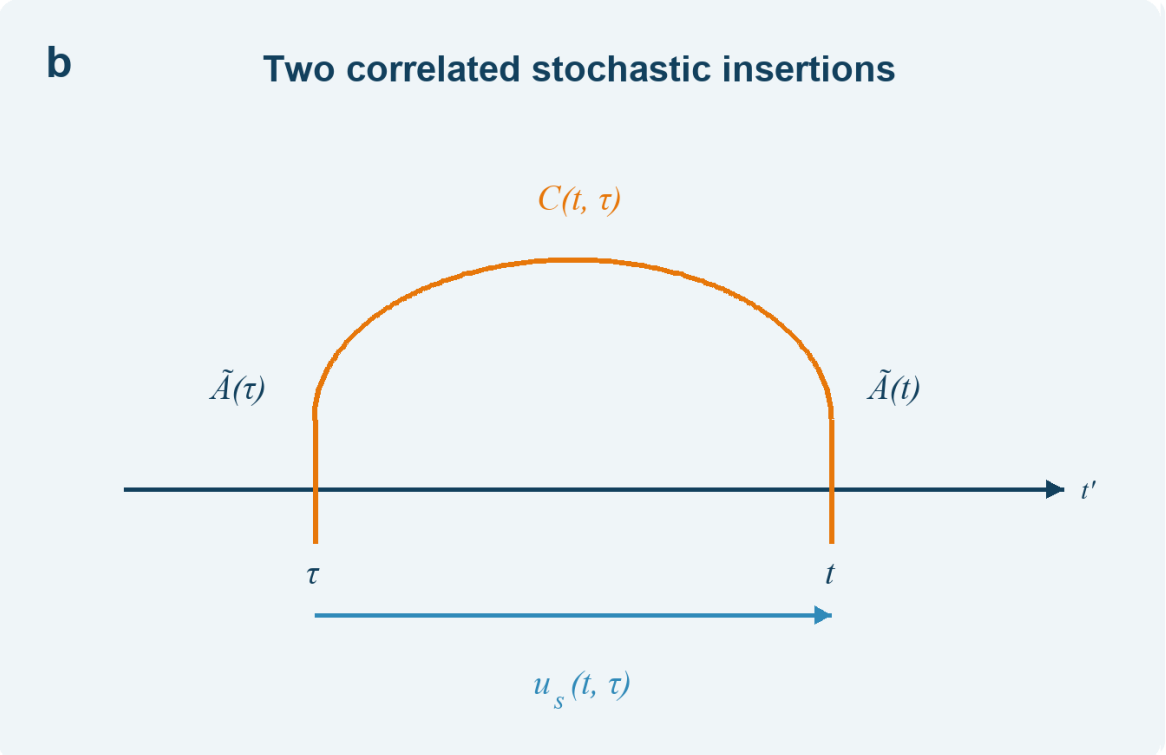

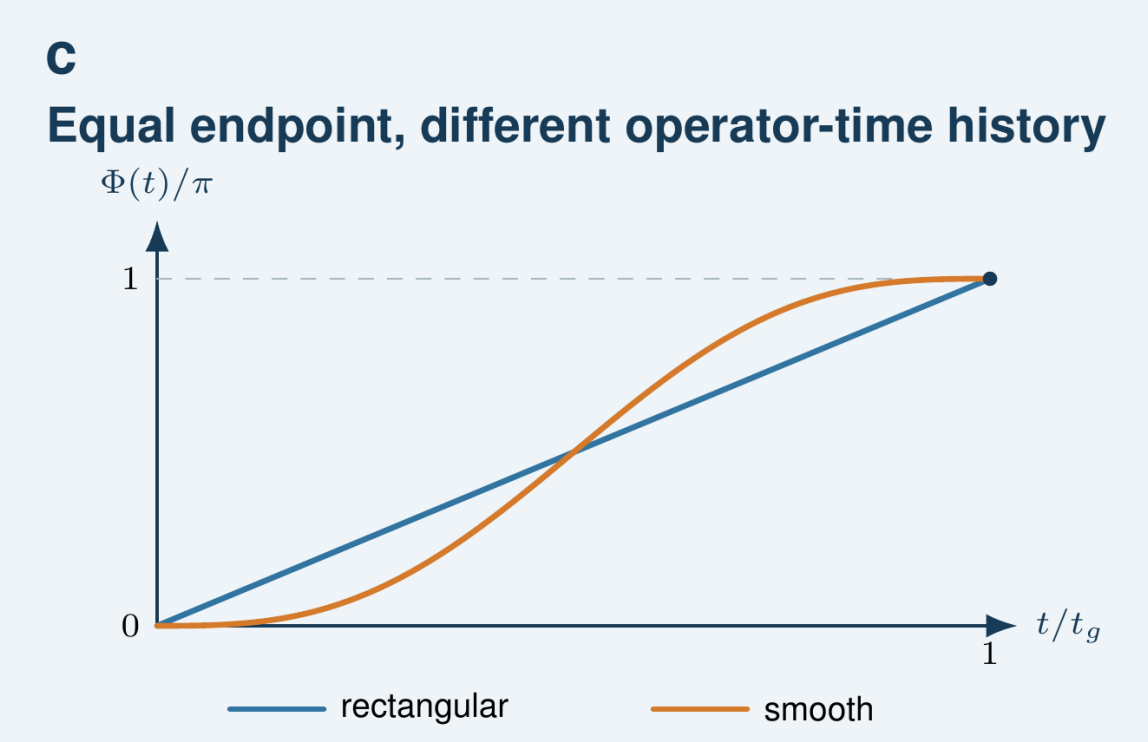

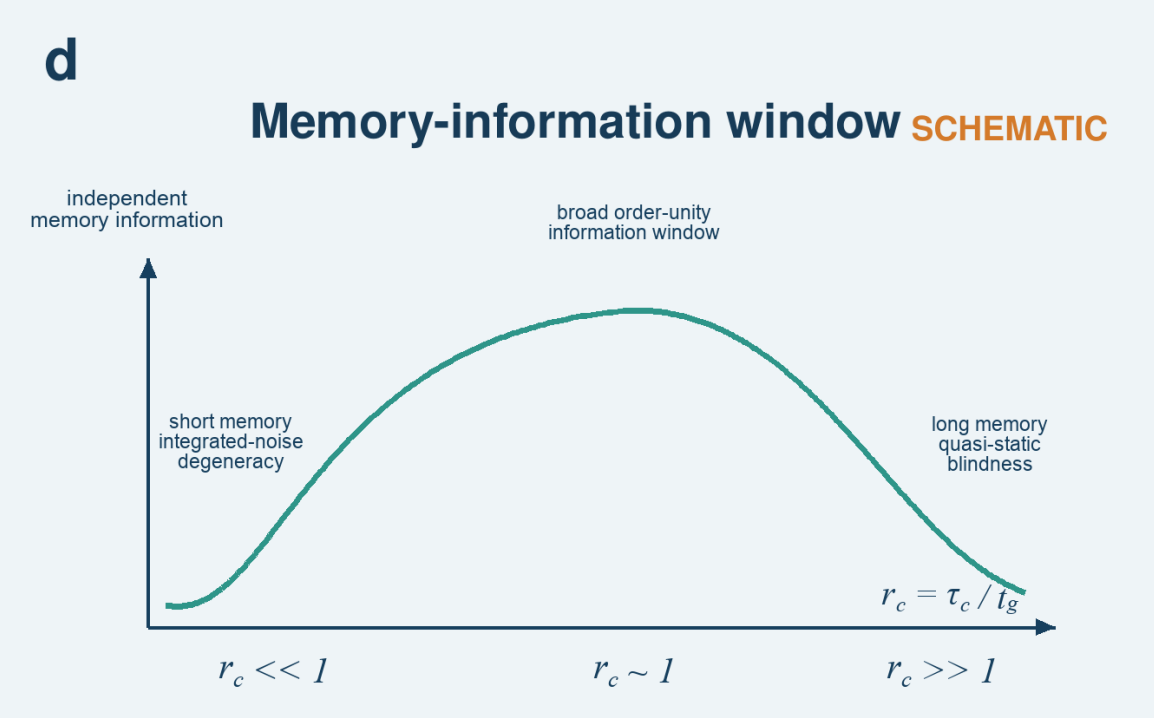

**Fig. 1 | Time-ordered control dressing of correlated stochastic interactions. a, A time-dependent coherent propagator dresses the stochastic coupling operator without assuming commuting Hamiltonians. b, Environmental correlation connects stochastic insertions at two times while coherent evolution fixes the operator orientation at each insertion. c, Equal-endpoint rectangular and smooth Xπ controls realize the same final unitary but different accumulated-phase histories; the panel compares temporal weighting along the same operator path, not different geometric paths. d, Short-memory integrated-noise degeneracy and long-memory quasi-static blindness bound a broad, order-unity memory-information window. Panel d is schematic and does not assert a universal maximum at correlation ratio unity.**

The environmental correlation function determines which pairs of stochastic interaction times remain statistically connected. The time-ordered coherent propagator independently determines the operator orientation associated with each of those times. The observable imprint of environmental memory therefore depends jointly on temporal correlation and coherent operator history. Equal-endpoint controls can carry different information because they traverse the same dressed-operator path at different rates. This mechanism motivates the memory-information window tested with exact stochastic maps in the subsequent figures.

## Finite correlation time imposes two information limits

Only after the general dressed theory is established do we specialize to stationary Ornstein–Uhlenbeck noise,13,14

$$C(t-\tau) \;=\; \sigma^2 \exp(-\frac{|t-\tau|}{\tau_c}) \tag{12}$$

and define the dimensionless noise strength and correlation ratio

$$\lambda \;=\; \sigma t_g, \quad r_c \;=\; \frac{\tau_c}{t_g} \tag{13}$$

The correlation ratio in Eq. (13) compares the environmental correlation time with the finite duration over which the quantum process probes the environment. For short memory, the correlation confines the two-time integral to a narrow layer, and the leading process correction scales as

$$\delta\mathcal{E} \propto \lambda^2 r_c \quad (r_c \ll 1) \tag{14}$$

In dimensional variables the leading combination is the noise variance multiplied by the correlation time. The logarithmic derivatives with respect to dimensionless strength and correlation ratio are therefore proportional at leading order, producing a rank-deficient parameter-information direction. Noise sensitivity does not vanish; independent correlation-time identifiability vanishes.

For long memory,

$$e^{-\frac{|u-v|}{r_c}} = 1 - \frac{|u-v|}{r_c} + \mathcal{O}(r_c^{-2}) \quad (r_c \gg 1) \tag{15}$$

so the leading process is quasi-static and correlation-time dependence enters through suppressed corrections,

$$\left\| \frac{\partial\mathcal{E}}{\partial \log r_c} \right\| = \mathcal{O}(r_c^{-1}) \tag{16}$$

Equation (16) is an asymptotic order statement. For the single-qubit rectangular control, a finite-range log–log fit at $r_c$ = 1, 3 and 10 gives a slope of approximately −0.69. The two-qubit exchange calculation shows the same qualitative suppression of correlation-time sensitivity with increasing $r_c$. The single-qubit finite-range slope is not identified with the asymptotic exponent because, at long memory, the exact correlation-time derivative approaches the Monte Carlo floor, i.e., the statistical resolution limit set by finite trajectory sampling. Below this level, further physical suppression of the derivative cannot be reliably distinguished from finite-sampling fluctuations.

From each complete exact process map $\mathcal{E}$, we construct its Pauli-transfer matrix (PTM), $R$, which represents the action of the quantum channel on a fixed Pauli-operator basis, and form its logarithmic-parameter Jacobian,

$$R_{ij} = \frac{1}{d}\mathrm{Tr}[P_i\, \mathcal{E}(P_j)], \quad J = [\frac{\partial\, \mathrm{vec}\, R}{\partial \log \lambda}, \frac{\partial\, \mathrm{vec}\, R}{\partial \log r_c}] \tag{17}$$

The correlation-time derivative measures absolute sensitivity, whereas the smallest singular value of J isolates the direction independent of the dominant noise-strength derivative. The smallest eigenvalue of $J^TJ$ is a Fisher-like full-process local information metric, not quantum Fisher information. The relation between this full-process metric and experimentally compressed gate- and PTM-level observables is examined in Supplementary Note IV and Supplementary Figs. S6–S9.

## Exact one- and two-qubit maps reveal a memory-information window

### Driven single qubit

For an X-axis coherent drive with transverse Z noise, the dressed stochastic operator is

$$\mathcal{Z}(t) = u_s{}^{\dagger}(t,0) Z u_s(t,0) = Z\cos\Phi(t) + Y\sin\Phi(t), \quad \Phi(t) = \int_0^t \Omega(v)\mathrm{d}v \tag{18}$$

Rectangular and smooth equal-area pulses implement the same Xπ endpoint but accumulate Φ(t) at different rates. A Z control with Z noise is the commuting null: the stochastic operator is not dynamically rotated.

### Two-qubit exchange

For local stochastic $Z_1$ coupling, isotropic exchange mixes the noise operator into a larger one- and two-body operator sector. The exchange Hamiltonian is nevertheless a special commuting-in-time coherent control,

$$H_s(t) = J(t)\,\boldsymbol{S}_1 \cdot \boldsymbol{S}_2, \quad u_s(t,\tau) = \exp[-i\Phi_J(t,\tau)\,\boldsymbol{S}_1 \cdot \boldsymbol{S}_2] \tag{19}$$

$$\Phi_J(t,\tau) = \int_\tau^t J(v)\mathrm{d}v, \quad [H_s(t_1), H_s(t_2)] = 0 \tag{20}$$

Thus the general framework retains time ordering, while this exchange specialization permits the exact ordinary exponential because its Hamiltonians commute at distinct times. The commuting null uses $\xi(t)Z_1Z_2$, for which exchange does not rotate the stochastic operator. The full dressed Pauli algebra remains in the Supplementary Information.

The resulting exact single- and two-qubit memory-information windows are shown in Fig. 2.

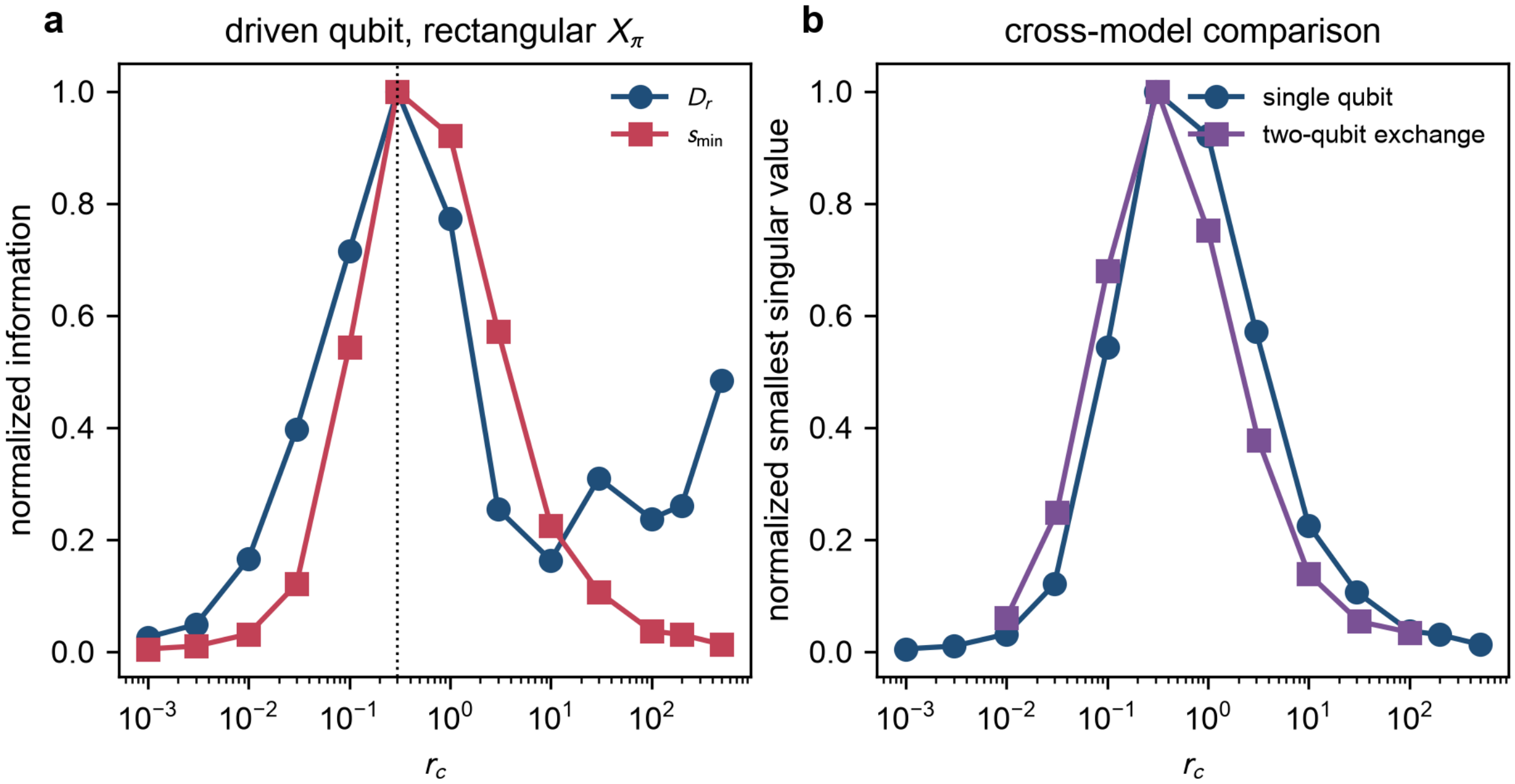

**Fig. 2 | Exact memory-information windows. a, Normalized correlation-time derivative and smallest PTM-Jacobian singular value for a rectangular single-qubit $X_\pi$ gate at λ=0.084. b, Peak-normalized smallest singular values for the single-qubit and full informative two-qubit maps. Raw values are not compared across dimensions.**

The rectangular driven-qubit window peaks at $r_c$≈0.3 with approximate half-maximum range 0.1–1; the smooth pulse broadens the useful interval to approximately 0.1–3. The two-qubit exchange map displays the same rise and fall. Thus “comparable timescales” denotes an order-unity interval, not an exact crossover at $r_c$=1.

### Long-memory suppression and pulse-history dependence

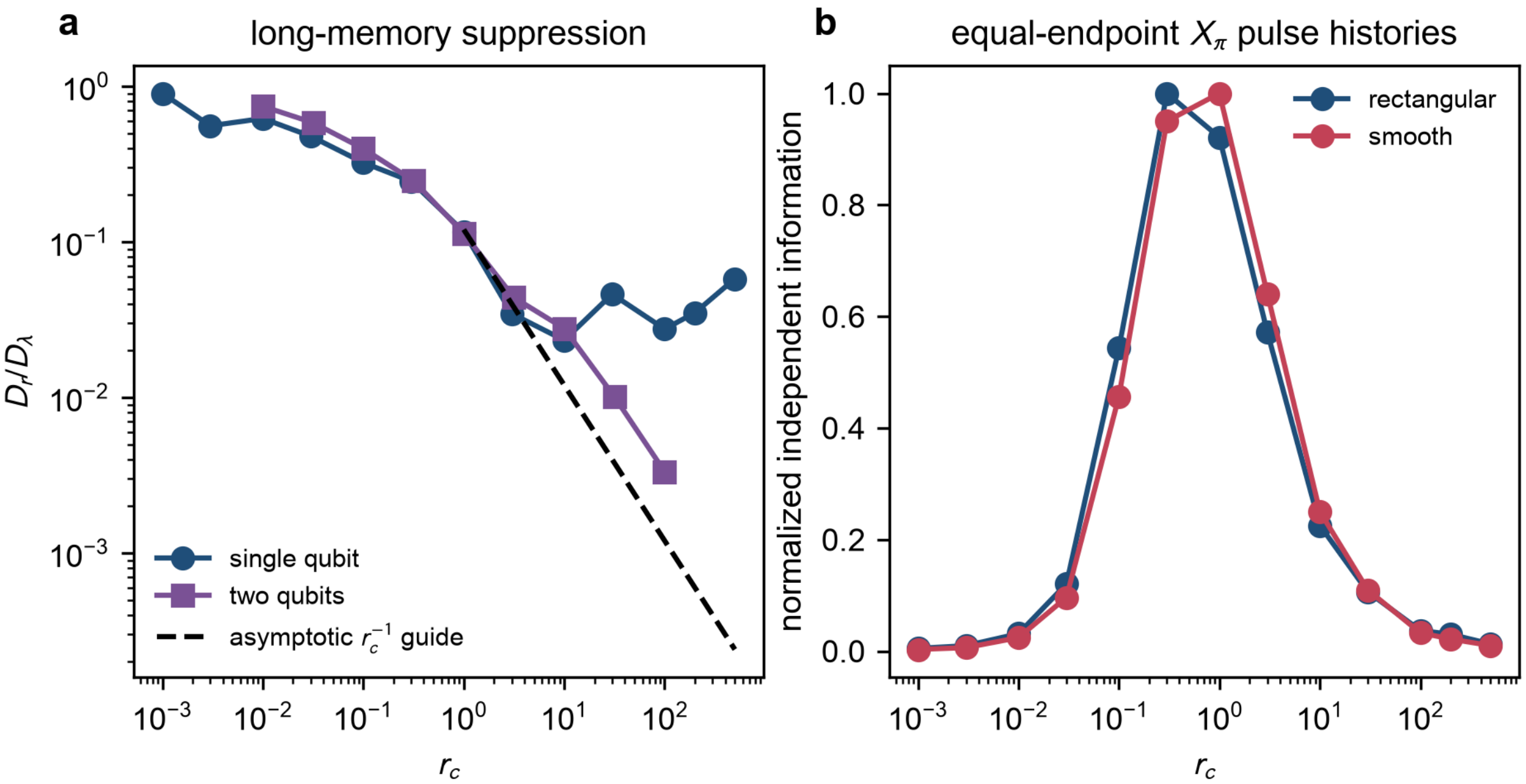

**Fig. 3 | Asymptotic loss and control-history dependence. a, Exact $D_r/D_\lambda$ ratios. A finite-range log–log fit for the single-qubit rectangular control at $r_c$ = 1, 3 and 10 gives a slope of approximately −0.69. The two-qubit exchange calculation shows the same qualitative long-memory suppression. The dashed $r_c^{-1}$ line is shown only as an asymptotic guide and is not a fit to the numerical data. b, Equal-endpoint rectangular and smooth $X_\pi$ gates encode different memory profiles.**

In the long-memory regime, correlation-time sensitivity decreases as the stochastic field becomes quasi-static over the gate interval. The single- and two-qubit calculations show this suppression; the finite-range single-qubit slope quoted in Fig. 3a is a numerical fit rather than an asymptotic exponent. Equal endpoint, duration and integrated area do not imply equal memory observability: the timing of the dressed-operator path weights the OU kernel differently. No universal smooth-pulse advantage is claimed.

### Noncommuting control opens an independent memory-information direction

The effect of commuting and noncommuting control on the independent memory-information direction is shown in Fig. 4.

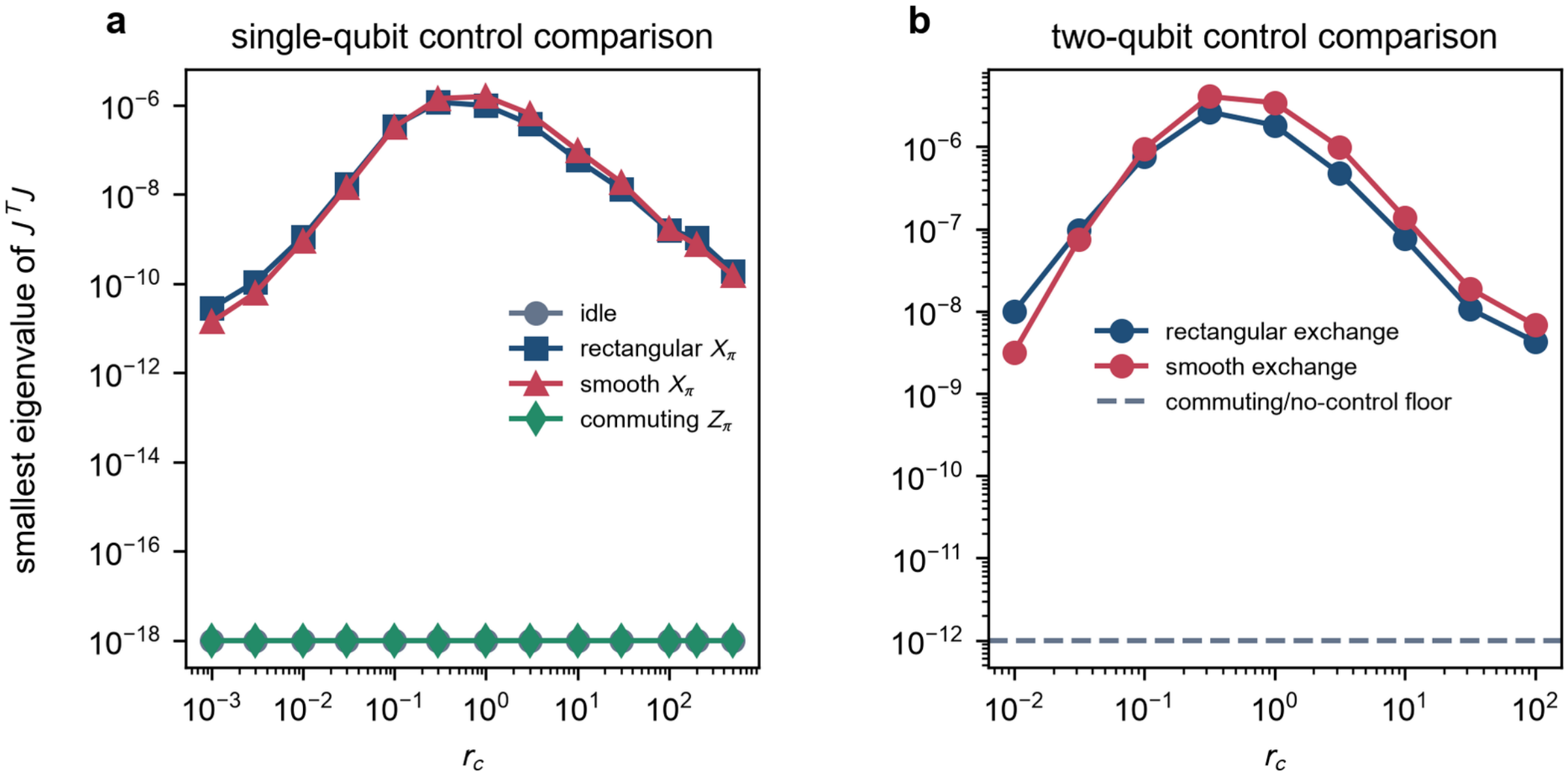

**Fig. 4 | Control-induced memory observability. a, Single-qubit smallest eigenvalue of $J^TJ$ for idle, noncommuting $X_\pi$ and commuting $Z_\pi$ controls. b, Two-qubit exchange controls with the commuting/no-control level at the numerical floor.**

Idle dephasing and commuting controls remain rank one because their error channels depend on a single accumulated phase-variance scalar. Noncommuting control opens a finite second direction with an intermediate maximum. Applying a pulse is not sufficient; it must rotate the stochastic operator between correlated events. Additional pulse-history and PTM-resolved null tests show the same collapse to numerical precision (Supplementary Note IV and Supplementary Fig. S10).

## The information window follows the gate duration

The gate-duration scaling and its direct exact-OU validation are shown in Fig. 5.

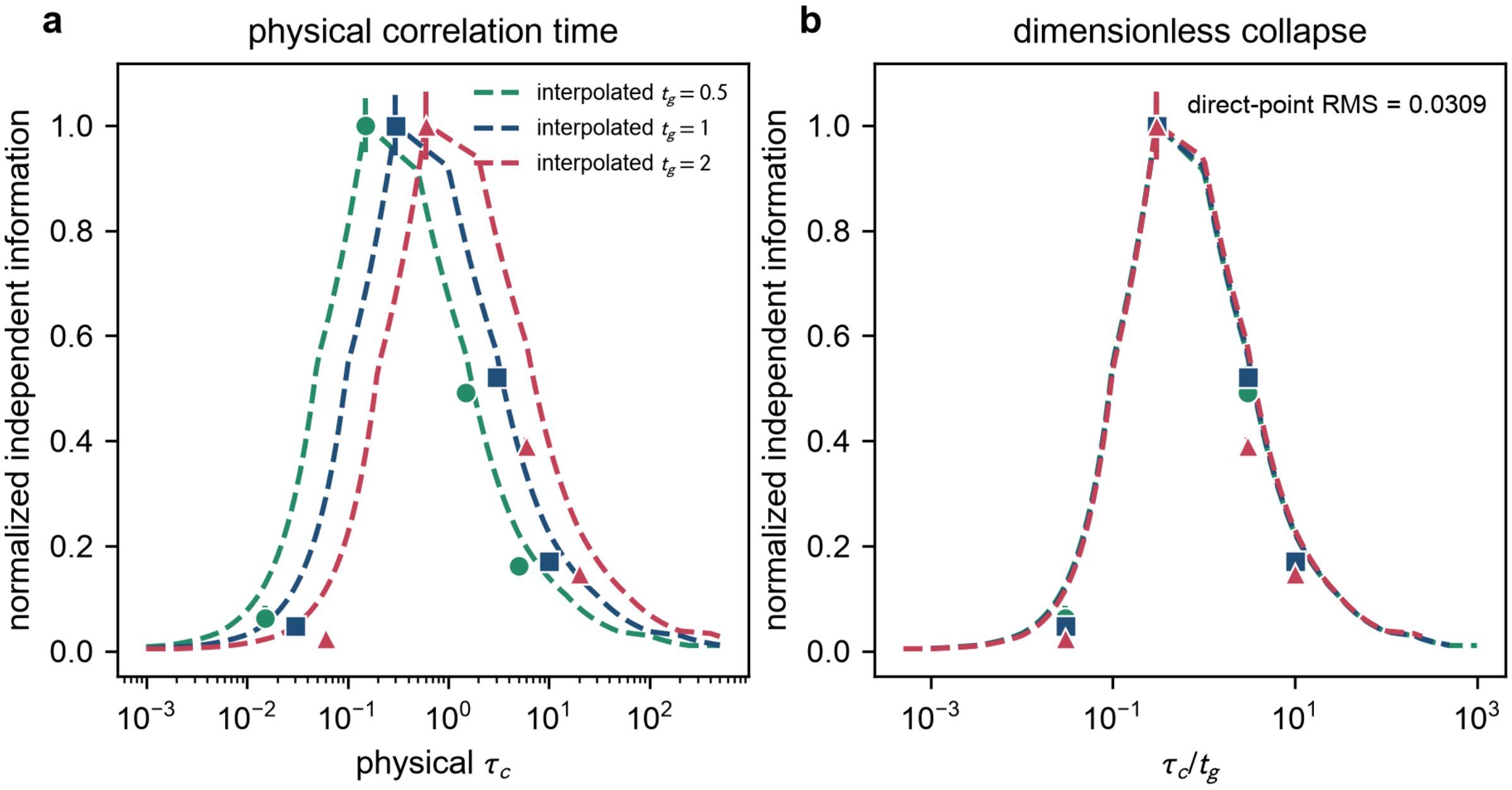

**Fig. 5 | Gate-duration scaling and direct exact-OU validation. a, Dashed curves are peak-normalized interpolants for $X\pi$ gates with $t_g$=0.5, 1 and 2; markers with Monte Carlo standard errors are new direct physical-parameter**

**simulations below, near and above the peak. b, The same data versus $\frac{\tau_c}{t_g}$. For each gate duration, the direct validation points and the interpolated curve were normalized by the peak of the corresponding interpolated information curve. The direct validation points give a normalized collapse RMS of 0.0309, whereas the interpolation-only curves give an internal RMS of 0.0046. The scaling is approximate, not an exact universal collapse.**

The physical peak shifts with gate duration, while directly simulated points show approximate duration collapse against $\frac{\tau_c}{t_g}$ with normalized RMS 0.0309. Twelve new points used $\sigma$=0.065, four correlation ratios per duration, 2,000 trajectories per point and centered logarithmic derivatives. Only three points agree pointwise with the coarse-grid interpolant within two Monte Carlo standard errors, and relative deviations range from 2% to 72%. These deviations test pointwise agreement of the absolute information magnitude and expose interpolation and sampling error. The normalized RMS collapse instead tests the shape shared by separately normalized duration curves; large local magnitude deviations can therefore coexist with a much smaller shape-collapse RMS. The interpolation-only internal RMS 0.0046 is retained solely as a smooth-curve diagnostic, whereas 0.0309 is the direct-validation result.

## Discussion

The results identify two complementary information ceilings and make a distinction that is essential for interpreting noisy gates: an operation can be strongly affected by noise yet contain little independent information about its correlation time. At short memory, finite control retains primarily $\sigma^2\tau_c$. At long memory, the field is effectively constant during the operation and cannot reveal how slowly it will change after the operation ends.

The two models differ in Hilbert-space dimension, gate, operator algebra and map size. Their normalized agreement is therefore nontrivial. In one qubit, $Z$ noise rotates through $Y$ and $Z$; under exchange, local noise mixes into two-body sectors. The common element is the coherent operator trajectory between stochastic insertions.

The supported scope remains stationary Gaussian OU noise. The exponential correlation fixes prefactors and crossover shape. Timescale matching and quasi-static blindness may extend to other finite-correlation environments, but random telegraph noise, nonstationarity, strong non-Gaussianity and arbitrary controls are not tested here.

Noncommuting control can create new process directions by rotating the stochastic operator between correlated events. The commuting nulls show that applying a pulse alone is insufficient. Finally, no measurement strategy can recover a parameter whose derivative has vanished from the underlying process. In that regime, the observation or control duration must change. The direct duration-scaling result connects this identifiability ceiling to an actionable design: shift the information window by shifting the gate timescale. This limitation persists even when the complete informative process map is retained: the full-PTM analysis becomes strongly ill-conditioned in the long-memory regime (Supplementary Note III and Supplementary Fig. S5).

Within stationary finite-correlation Gaussian OU noise and the one- and two-qubit systems studied here, timescale matching is a control principle for probing non-Markovian dynamics.

## Methods

### Exact stationary Ornstein–Uhlenbeck sampling

The exact transition is $\xi_{n+1} = a\xi_n + \sigma\sqrt{1-a^2}\eta_n$, $a = \exp(\frac{-\Delta t}{\tau_c})$. With stationary Gaussian initialization and independent unit-normal innovations, this transition is exact at the discrete sample times for a stationary OU process.

### Quantum trajectory propagation

Each realization is propagated with the time-dependent control Hamiltonian and paired with −ξ(t). Antithetic pairs cancel leading odd sampling fluctuations without changing the ensemble. Single-qubit steps use exact SU(2) Pauli exponentials; two-qubit steps use the complete trajectory unitary.

### Pauli-transfer-matrix construction

Trajectory unitaries are converted to channel superoperators before ensemble averaging. Two-qubit trajectories generate complete 16×16 PTMs; single-qubit trajectories generate complete 4×4 PTMs. Exact maps are convex averages of unitary channels and are therefore completely positive and trace preserving (CPTP) by construction.15 The second-order time-convolutionless (TCL2) approximation is used only to organize the two-time analytical structure and for supplementary benchmarks, not for the central numerical conclusions.

### Process sensitivity and identifiability metrics

For dimension $d$, $R_{ij} = \frac{1}{d} Tr[P_i\, \mathcal{E}(P_j)]$ in a fixed Pauli order. Identity-fixed entries are excluded from $J$. Frobenius norms are evaluated with the same normalized basis within each system; cross-dimensional comparisons use peak-normalized curves, not raw singular values.

### Single-qubit calculations

The grid used λ={0.03,0.05,0.065,0.084,0.10,0.15} and $r_c$={0.001,0.003,0.01,0.03,0.1,0.3,1,3,10,30,100,200,500}, with 500 antithetic pairs and adaptive resolution up to 4,096 time steps.

### Two-qubit exchange calculations

The exact library used seven strengths, nine correlation ratios, three exchange histories and common random numbers. Hidden tests used 4,000 trajectories and 256 steps. Complete settings and seeds accompany the manuscript.

Complete numerical parameter grids, trajectory settings and representative convergence values are provided in Supplementary Tables S1–S3.

### Numerical convergence and uncertainty

Logarithmic derivatives use centered or grid-consistent finite differences with shared random numbers wherever possible. Log-step values 0.03 and 0.06 agreed within 0.05%. Sampling-relative spreads for $D_r$ were 1.5%, 4.3% and 7.3% at $r_c$=0.01, 0.3 and 30, respectively. The increase at long memory is expected as the derivative approaches the Monte Carlo floor. Validation of the stationary OU sampler, propagation convergence, exact-OU/TCL2 comparisons and complete-positivity diagnostics are provided in Supplementary Note III and Supplementary Figs. S1–S4.

### Gate-duration scaling and direct validation

At fixed physical $\sigma$, $\lambda=\sigma t_g$ and $r_c=\frac{\tau_c}{t_g}$. Exact-library interpolants were evaluated for $t_g$={0.5,1,2}. The interpolation-only peak-normalized curves give RMS 0.0046, which is retained only as an internal smooth-curve metric.

Four direct correlation ratios {0.03,0.3,3,10} were simulated for each duration, giving twelve physical points. Each point used ten independent batches of 100 antithetic pairs, 2,000 trajectories in total, centered logarithmic step 0.04 and at least 256 time steps. Batch variation supplied the Monte Carlo standard error. Direct information values were compared with the interpolated curves before normalization; absolute, relative and standardized deviations are recorded. For each duration, the direct validation points and the interpolated curve were normalized by the peak of the corresponding interpolated information curve; the direct-point collapse RMS was 0.0309.

### Statistics

No p values are reported. Uncertainty is assessed through antithetic Monte Carlo sampling, resolution tests and finite-difference-step convergence. No parameter points were removed after inspection.

## Use of generative AI tools

OpenAI ChatGPT and Codex were used to assist with numerical-code development and debugging, equation and figure formatting, and language editing. Anthropic Claude was additionally used for cross-checking consistency among the manuscript, Supplementary Information, numerical source data, and reproducibility package. All mathematical derivations, numerical algorithms, simulation parameters, numerical outputs, figures, and scientific interpretations were independently checked and validated by the author. The AI tools were not used as autonomous sources of research data or to determine the scientific conclusions of the study.

## Data availability

The numerical source data supporting the findings of this study, including exact process maps, PTM derivatives, information-window metrics, pulse-history comparisons, commuting-control tests, convergence analyses and gate-duration scaling data, are available in Zenodo at https://doi.org/10.5281/zenodo.22680305

## Code availability

The custom code used for stationary Ornstein–Uhlenbeck sampling, stochastic trajectory propagation, Pauli-transfer-matrix construction, process-sensitivity analysis and figure generation is available in the same Zenodo repository at https://doi.org/10.5281/zenodo.22680305

## Acknowledgements

This work was supported by Korea National Research Foundation (NRF) grant No. NRF-2026-25612745, ICT R&D program of IITP-2023- 2021-0-01810, RS-2023-00225385, AFOSR grant FA2386-22-1-4052, Cleveland Clinic Quantum Innovation Catalyzer Program and Amazon Web Services. We acknowledge the use of IBM Quantum services for this work. The views expressed are those of the author, and do not reflect the official policy or position of IBM or the IBM Quantum team. This work was also supported by the National Quantum Laboratory at Maryland (QLab). We acknowledge IonQ for providing quantum computing resources for this work. The views expressed are those of the author, and do not reflect the official policy or position of IonQ.

## Author contributions

D.A. conceived the study, developed the theoretical framework, supervised the numerical analysis, interpreted the results, and wrote the manuscript.

## Competing interests

D.A. is the founder of Singularity Quantum Inc. and holds an equity interest in the company. The author declares no other competing interests.

# Supplementary Information for
# A memory-information window in finite-duration quantum control

Doyeol Ahn[1,2 *]

[1] Department of Electrical and Computer Engineering,
University of Seoul, 163 Seoulsiripdae-ro, Tongdaimoon-gu, Seoul 02504, Republic of Korea

[2] Singularity Quantum Inc
895 Dove Street Second Floor, Newport Beach, CA 92660, USA

*Corresponding author: dahn@uos.ac.kr

**I. Liouville-Space Propagation under a Time-Dependent System Hamiltonian**

**Definition of the Liouville superoperator**

For an arbitrary operator, we define the time-dependent Liouville superoperator by

$$\mathcal{L}_s(t)X = [H_s(t), X] \tag{S1}$$

This superoperator is the adjoint representation of the Hamiltonian, and its action on operator space is the commutator with the instantaneous system Hamiltonian. The definition does not require Hamiltonians at distinct times to commute.

**Static-Hamiltonian identity**

For a time-independent Hamiltonian, we introduce the conjugated operator

$$F(\lambda) = e^{\lambda H} X e^{-\lambda H}$$

By differentiating this expression, we obtain the operator initial-value problem

$$\frac{dF}{d\lambda} = L_H F, F(0) = X$$

Therefore, from the uniqueness of the operator initial-value problem, we obtain the static adjoint-action identity

$$e^{\lambda \mathcal{L}_H} X = e^{\lambda H} X e^{-\lambda H} \tag{S2}$$

The same result can be written as the Hadamard expansion

$$e^{\lambda \mathcal{L}_H} X = X + \lambda [H, X] + \frac{\lambda^2}{2!} [H, [H, X]] + \cdots$$

**Lemma: time-dependent Hamiltonian**

**Lemma.** Let the system propagator and corresponding Liouville-space propagator be

$$u_s(t,\tau) = \mathcal{T}exp\left[-i\int_\tau^t dv H_s(v)\right], \mathcal{U}_s(t,\tau) = \mathcal{T}exp\left[-i\int_\tau^t dv \mathcal{L}_s(v)\right]. \tag{S3}$$

Then, for every operator,

$$\mathcal{U}_s(t,\tau)X = u_s(t,\tau)Xu_s^\dagger(t,\tau). \tag{S4}$$

**Proof.** The propagator and its adjoint satisfy

$$\frac{\partial}{\partial t}u_s(t,\tau) = -iH_s(t)u_s(t,\tau), \frac{\partial}{\partial t}u_s^\dagger(t,\tau) = +iu_s^\dagger(t,\tau)H_s(t)$$

We define the propagated operator by conjugation. By applying the product rule and the two differential equations above, we obtain

$$\begin{aligned}\frac{dF(t)}{dt} &= -iH_s(t)F(t) + iF(t)H_s(t)\\ &= -i[H_s(t),F(t)] = -i\mathcal{L}_s(t)F(t), F(\tau) = X.\end{aligned} \tag{S5}$$

The Liouville-space expression in Eq. (S4) satisfies the same initial-value problem. The uniqueness of the linear operator differential equation therefore proves Eq. (S4).

**Why time ordering is essential**

In general, we do not assume that Hamiltonians at distinct times commute. Therefore,

$$u_s(t,\tau) \neq exp\left[-i\int_\tau^t dvH_s(v)\right] \textit{ in general.} \tag{S6}$$

Thus, the time-ordering operator must be retained. Equivalently, the adjoint action is

$$\left[\mathcal{T}e^{-i\int_\tau^t dvH_s(v)}\right]X\left[\mathcal{T}^C e^{+i\int_\tau^t dvH_s(v)}\right].$$

Here, $T_C$ denotes anti-time ordering, and the second factor is the adjoint system propagator.

**Connection to the control-dressed interaction**

For a system coupling operator indexed by $\alpha$, we define the control-dressed operator by

$$\tilde{A}_\alpha(t) = u_s^\dagger(t,0)A_\alpha u_s(t,0) = \mathcal{U}_s(0,t)A_\alpha = \mathcal{U}_s(t,0)^{-1}A_\alpha. \tag{S7}$$

This expression is the inverse Liouville propagator acting on the coupling operator. It is therefore the adjoint action generated by the system Hamiltonian in the interaction, or control-dressed, frame.

This identity gives the dressed system operators in the stochastic interaction Hamiltonian. Their complete time-ordered control dependence then enters the TCL2 expansion and the non-Markovian memory kernel without an additional physical assumption [1].

**Corollary: commuting Hamiltonians**

**Corollary.** If the Hamiltonian commutes with itself at all pairs of times,

$$[H_s(t_1), H_s(t_2)] = 0 \textit{ for all} t_1, t_2$$

the time-ordering operator is not required, and

$$u_s(t,\tau) = exp\left[-i\int_\tau^t dvH_s(v)\right]. \tag{S8}$$

Therefore,

$$\mathcal{U}_s(t,\tau)X = e^{-i\int_\tau^t dvH_s(v)}Xe^{+i\int_\tau^t dvH_s(v)}. \tag{S9}$$

Thus, the ordinary-exponential form in the original lemma applies when the Hamiltonians commute, rather than for an arbitrary time-dependent Hamiltonian.

**Exchange-control specialization**

For the exchange-control Hamiltonian

$$H_s(t) = J(t)\boldsymbol{S}_1 \cdot \boldsymbol{S}_2$$

the operator part is time independent. We therefore obtain

$$[H_s(t_1), H_s(t_2)] = J(t_1)J(t_2)[\boldsymbol{S}_1 \cdot \boldsymbol{S}_2, \boldsymbol{S}_1 \cdot \boldsymbol{S}_2] = 0. \tag{S10}$$

We define the accumulated exchange phase by

$$\Phi(t,\tau) = \int_\tau^t dv J(v).$$

Then, the system propagator is

$$u_s(t,\tau) = exp[-i\Phi(t,\tau)\boldsymbol{S}_1 \cdot \boldsymbol{S}_2]$$

and the corresponding Liouville-space action becomes

$$\mathcal{U}_s(t,\tau)X = e^{-i\Phi(t,\tau)\boldsymbol{S}_1\cdot\boldsymbol{S}_2} X e^{+i\Phi(t,\tau)\boldsymbol{S}_1\cdot\boldsymbol{S}_2} \quad . \tag{S11}$$

The general theory allows noncommuting system Hamiltonians at different times and therefore retains time ordering. The exchange-control model is a special commuting case, for which the propagator reduces exactly to an ordinary exponential.

## II. Time-convolutionless evolution of a non-Markovian reduced-density operator

The total Hamiltonian for an open two-state system is given by [2–6]

$$H_T(t) = H_s(t) + H_i(t) \tag{S12}$$

where $H_s(t)$ is the system Hamiltonian of the two-state system, and $H_i(t)$ describes its interaction with the stochastic reservoir.

The equation of motion for the total density operator $\rho_T(t)$ is given by the stochastic Liouville equation [2,7]

$$\frac{d\rho_T(t)}{dt} = -i[H_T(t), \rho_T(t)] = -i\mathcal{L}_T(t)\rho_T(t), \tag{S13}$$

where

$$\mathcal{L}_T(t) = \mathcal{L}_s(t) + \mathcal{L}_i(t) \tag{S14}$$

is the Liouville superoperator corresponding to the Hamiltonian. We use units in which $\hbar = 1$. To obtain an equation for the system alone, we introduce time-independent projection operators that eliminate the reservoir degrees of freedom. We define the projection operators $P$ and $Q$ by [2–5]

$$PX = \langle X \rangle_i \tag{S15}$$

for any dynamical variable $X$. Here, $\langle\cdots\rangle$_$i$ denotes the average over the stochastic process $H_i(t)$. The projection operators satisfy

$$P^2 = P, Q^2 = Q, PQ = QP = 0. \tag{S16}$$

The system information is contained in the reduced density operator

$$\rho(t) = P\rho_T(t). \tag{S17}$$

By multiplying Eq. (S13) by $P$ and $Q$ from the left, we obtain the coupled equations for $P\rho_T(t)$ and $Q\rho_T(t)$:

$$\frac{d}{dt}P\rho_T(t) = -iP\mathcal{L}_T(t)P\rho_T(t) - iP\mathcal{L}_T(t)Q\rho_T(t), \tag{S18a}$$

and

$$\frac{d}{dt}Q\rho_T(t) = -iQ\mathcal{L}_T(t)Q\rho_T(t) - iQ\mathcal{L}_T(t)P\rho_T(t), \tag{S18b}$$

where Eq. (S13) has been written as

$$\frac{d}{dt}\rho_T(t) = -i\mathcal{L}_T(t)\rho_T(T) = -i\mathcal{L}_T(t)(P+Q)\rho_T(t).$$

We assume that the interaction is turned on at $t = 0$ and that the total initial state is $\rho_T(0)$. We also assume that $Q\rho_T(0) = 0$. Under these conditions, the formal solution of Eq. (S18b) is

$$\begin{aligned} Q\rho_T(t) &= -i\int_0^t d\tau H(t,\tau)Q\mathcal{L}_T(\tau)P\rho_T(\tau) + H(t,0)Q\rho_T(0) \\ &= -i\int_0^t d\tau H(t,\tau)Q\mathcal{L}_T(\tau)PG(t,\tau)\rho_T(t) \\ &= -i\int_0^t d\tau H(t,\tau)Q\mathcal{L}_T(\tau)PG(t,\tau)(P+Q)\rho_T(t), \end{aligned} \tag{S19}$$

where the projected propagator $H(t,\tau)$ is defined by

$$H(t,\tau) = \mathcal{T}exp[-i\int_\tau^t dsQ\mathcal{L}_T(s)Q]. \tag{S20}$$

Here, $T$ denotes the time-ordering operator. We also introduce the anti-time evolution operator $G(t,\tau)$, defined by

$$G(t,\tau) = \mathcal{T}^C exp[i\int_\tau^t ds\mathcal{L}_T(s)]. \tag{S21}$$

which satisfies $\rho_T(\tau) = G(t,\tau)\rho_T(t)$. (S22)

It follows from Eqs. (S20) and (S21) [2,4] that

$$G(t,\tau)G(s,t) = G(s,\tau), H(t,\tau)H(\tau,s) = H(t,s). \tag{S23}$$

Here, $T_C$ denotes the anti-time-ordering operator.

Eq. (S19) can be rewritten as

$$[1 + i\int_0^t d\tau H(t,\tau)Q\mathcal{L}_T(\tau)PG(t,\tau)]Q\rho_T(t) = -i\int_0^t d\tau H(t,\tau)Q\mathcal{L}_T(\tau)PG(t,\tau)P\rho_T(t). \tag{S24}$$

We introduce the superoperators $g(t)$ and $\theta(t)$, defined by

$$g(t) = 1 + i\int_0^t d\tau H(t,\tau)Q\mathcal{L}_T(\tau)PG(t,\tau) \equiv \theta^{-1}(t) \tag{S25}$$

Then, we obtain

$$Q\rho_T(t) = [\theta(t) - 1]P\rho_T(t) \tag{S26}$$

and

$$\frac{d}{dt}P\rho_T(t) + iP\mathcal{L}_T(t)P\rho_T(t) = -iP\mathcal{L}_T(t)[\theta(t)-1]P\rho_T(t). \quad \text{(S27)}$$

After some mathematical manipulations, we obtain the following formal solution of Eq. (S27):

$$\begin{aligned} P\rho_T(t) &= \mathcal{U}(t,0)P\rho_T(0) - i\int_0^t ds\mathcal{U}(t,s)P\mathcal{L}_T(s)[\theta(s)-1]P\rho_T(s) \\ &= \mathcal{U}(t,0)P\rho_T(0) - i\int_0^t ds\mathcal{U}(t,s)P\mathcal{L}_T(s)[\theta(s)-1]PG(t,s)\rho_T(t) \\ &= \mathcal{U}(t,0)P\rho_T(0) - i\int_0^t ds\mathcal{U}(t,s)P\mathcal{L}_T(s)[\theta(s)-1]PG(t,s)(P+Q)\rho_T(t) \\ &= \mathcal{U}(t,0)P\rho_T(0) - i\int_0^t ds\mathcal{U}(t,s)P\mathcal{L}_T(s)[\theta(s)-1]PG(t,s)P\rho_T(t) \\ &\quad -i\int_0^t ds\mathcal{U}(t,s)P\mathcal{L}_T(s)[\theta(s)-1]PG(t,s)Q\rho_T(t) \\ &= \mathcal{U}(t,0)P\rho_T(0) - i\int_0^t ds\mathcal{U}(t,s)P\mathcal{L}_T(s)[\theta(s)-1]PG(t,s)\theta(t)P\rho_T(t) \end{aligned}, \quad \text{(S28)}$$

where

$$\mathcal{U}(t,\tau) = \mathcal{T}exp[-i\int_\tau^t dsP\mathcal{L}_T(s)P] \quad \text{(S29)}$$

This is the projected propagator of the total system.

From $P\mathcal{L}_T(t)P = P(\mathcal{L}_s(t)+\mathcal{L}_i(t))P = P\mathcal{L}_s(t)P = \mathcal{L}_s(t)P^2 = \mathcal{L}_s(t)P$ , we obtain

$$\begin{aligned} \mathcal{U}(t,s)P &= \mathcal{T}exp[-i\int_s^t d\tau P\mathcal{L}_T(\tau)P]P \\ &= \mathcal{T}exp[-i\int_s^t d\tau\mathcal{L}_s(\tau)P]P \\ &= \mathcal{T}exp[-i\int_s^t d\tau\mathcal{L}_s(\tau)P] = \mathcal{U}_s(t,s)P. \end{aligned} \quad \text{(S30)}$$

Here, $U_s(t,s)$ is the system propagator.

From Eqs. (S15) and (S17), we obtain

$$\begin{aligned} &\mathcal{U}(t,s)P\mathcal{L}_T(s)[\theta(s)-1]PG(t,s)\theta(t)P\rho_T(t) \\ &= \mathcal{U}_s(t,s)\langle\mathcal{L}_i(s)[\theta(s)-1]\rangle_i\langle G(t,s)\theta(t)\rangle_i\rho(t). \end{aligned} \quad \text{(S31)}$$

By substituting Eq. (S31) into Eq. (S28) and using Eqs. (S29) and (S30), we obtain

$$\rho(t) = \mathcal{U}_s(t,0)\rho(0) - i\int_0^t ds\mathcal{U}_s(t,s)\langle\mathcal{L}_i(s)[\theta(s)-1]\rangle_i\langle G(t,s)\theta(t)\rangle_i\rho(t), \quad \text{(S32)}$$

or

$$\begin{aligned} &(1 + i\int_0^t ds\mathcal{U}_s(t,s)\langle\mathcal{L}_i(s)[\theta(s)-1]\rangle_i\langle G(t,s)\theta(t)\rangle_i)\rho(t) \\ &= \mathcal{U}_s(t,0)\rho(0). \end{aligned} \quad \text{(S33)}$$

We define $W(t)$ by

$$\mathcal{W}(t) = 1 + i\int_0^t ds\mathcal{U}_s(t,s)\langle\mathcal{L}_i(s)[\theta(s)-1]\rangle_i\langle G(t,s)\theta(t)\rangle_i, \quad \text{(S34)}$$

Then, the reduced density operator evolves as

$$\rho(t) = \mathcal{W}^{-1}(t)\mathcal{U}_s(t,0)\rho(0) = \mathcal{V}(t)\rho(0), \quad \text{(S35)}$$

where the superoperator $\mathcal{V}(t)$ for the evolution of the reduced density operator is defined by $\mathcal{V}(t) = \mathcal{W}^{-1}(t)\mathcal{U}_s(t,0)$. Within the Born approximation, we have

$$\mathcal{V}^{(2)}(t) = [1 - \int_0^t ds\int_0^s d\tau\mathcal{U}_s(t,s)\langle\mathcal{L}_i(s)\mathcal{U}_s(s,\tau)\mathcal{L}_i(\tau)\mathcal{U}_s^{-1}(s,\tau)\rangle_i\mathcal{U}_s^{-1}(t,s)]\mathcal{U}_s(t,0),$$

(S36)

where

$$\mathcal{U}_s(t,s) = \mathcal{T}exp[-i\int_s^t d\tau \mathcal{L}_s(\tau)]. \tag{S37}$$

A detailed derivation of Eq. (S36) is given below.

We start again from Eq. (S34).

$$\mathcal{W}(t) = 1 + i\int_0^t ds \mathcal{U}_s(t,s)\langle \mathcal{L}_i(s)[\theta(s)-1]\rangle_i \langle G(t,s)\theta(t)\rangle_i.$$

We define

$$\begin{aligned} \Sigma(t) &= 1-\theta^{-1}(t) \\ &= -i\int_0^t d\tau H(t,\tau)Q\mathcal{L}_T(\tau)PG(t,\tau)' \end{aligned} \tag{S38}$$

Then,

$$\begin{aligned} \theta^{-1}(t) &= 1-\Sigma(t), \\ \theta(t) &= \frac{1}{1-\Sigma(t)}, \\ \theta(t)-1 &= \frac{\Sigma(t)}{1-\Sigma(t)}. \end{aligned} \tag{S39}$$

Also, we have

$$\begin{aligned} P \quad & \mathcal{L}_T(t)[\theta(t)-1] \\ &= P(\mathcal{L}_s(t)+\mathcal{L}_i(t))\Sigma(t)[1-\Sigma(t)]^{-1}. \end{aligned}$$

The following identities permit the projected propagator to be expressed in the interaction-picture form used in Eq. (S40):

$$H(t,\tau) = V(t)\, S(t,\tau)\, V^{\dagger}(\tau), \quad V(t) = P + Q\, U_s(t)\, Q$$

$$H(t,\tau)\, Q = U_s(t)\, S(t,\tau)\, U_s^{-1}(\tau)\, Q$$

The explicit expression for $\Sigma(t)$ is

$$\begin{aligned} \Sigma(t) &= -i\int_0^t d\tau H(t,\tau)Q\mathcal{L}_T(\tau)PG(t,\tau) \\ &= -i\int_0^t d\tau H(t,\tau)Q\mathcal{L}_i(\tau)PG(t,\tau) \\ &= -i\int_0^t d\tau \mathcal{U}_s(t)S(t,\tau)\mathcal{U}_s^{-1}(\tau)Q\mathcal{L}_i(\tau)P\mathcal{U}_s(\tau)R(t,\tau)\mathcal{U}_s^{-1}(t), \end{aligned} \tag{S40}$$

where

$$R(t,\tau) = \mathcal{T}^C exp[i\int_0^t ds \mathcal{U}_s^{-1}(s)\mathcal{L}_i(s)\mathcal{U}_s(s)] \tag{S41}$$

is the evolution operator in the interaction picture, and

$$S(t,\tau) = \mathcal{T}exp[-i\int_\tau^t ds Q\mathcal{U}_s^{-1}(s)\mathcal{L}_i(s)\mathcal{U}_s(s)Q] \tag{S42}$$

is the projected propagator in the interaction picture. Then, Eq. (S34) can be written as

$$\begin{aligned} \mathcal{W}(t) = 1 + \quad & i\int_0^t ds \mathcal{U}_s(t,s)\langle \mathcal{L}_i(s)\Sigma(s)[1-\Sigma(s)]^{-1}\rangle_i \\ & \times \langle \mathcal{U}_s(s)R(t,s)\mathcal{U}_s^{-1}(t)[1-\Sigma(t)]^{-1}\rangle_i. \end{aligned} \tag{S43}$$

Within the Born approximation, Eq. (S43) becomes

$$\mathcal{W}^{(2)}(t) = 1 + i\int_0^t ds \mathcal{U}_s(t,s)\langle \mathcal{L}_i(s)\Sigma^{(1)}(s)\rangle_i \times \langle \mathcal{U}_s(s)\mathcal{U}_s^{-1}(t)\rangle_i. \tag{S44}$$

where

$$\begin{aligned}\Sigma^{(1)}(s) &= -i\int_0^s d\tau \mathcal{U}_s(s)\mathcal{U}_s^{-1}(\tau)Q\mathcal{L}_i(\tau)P\mathcal{U}_s(\tau)\mathcal{U}_s^{-1}(s) \\ &= -i\int_0^s d\tau \mathcal{U}_s(s,\tau)\mathcal{L}_i(\tau)\mathcal{U}_s^{-1}(s,\tau)\end{aligned} \tag{S45}$$

where we use the ansatz $PL_iP = 0$. From Eqs. (S44) and (S45), we obtain

$$\begin{aligned}[\mathcal{W}^{(2)}]^{-1}(t) &= 1 - i\int_0^t ds \mathcal{U}_s(t,s)\langle \mathcal{L}_i(s)\Sigma^{(1)}(s)\rangle_i \mathcal{U}_s^{-1}(t,s) \\ &= 1 - \int_0^t ds \int_0^s d\tau \mathcal{U}_s(t,s)\langle \mathcal{L}_i(s)\mathcal{U}_s(s,\tau)L_i(\tau)\mathcal{U}_s^{-1}(s,\tau)\rangle_i \mathcal{U}_s^{-1}(t,s).\end{aligned} \tag{S46}$$

By substituting Eq. (S46) into $V^{(2)}(t) = [W^{(2)}]^{-1}(t)U_s(t,0)$, we obtain Eq. (S36).

After some mathematical manipulations, $V^{(2)}(t)\rho(0)$ becomes

$$\begin{aligned}\rho(t) &= [\mathcal{W}^{(2)}]^{-1}(t)\mathcal{U}_s(t,0)\rho(0) \\ &= \mathcal{V}^{(2)}(t)\rho(0) \\ &= \mathcal{U}_s(t)\rho(0) \\ &-\int_0^t ds\int_0^s d\tau \mathcal{U}_s(t,s)\langle \mathcal{L}_i(s)\mathcal{U}_s(s,\tau)\mathcal{L}_i(\tau)\mathcal{U}_s^{-1}(s,\tau)\rangle_i \mathcal{U}_s^{-1}(t,s)\mathcal{U}_s(t,0)\rho(0).\end{aligned} \tag{S47}$$

We define $B$ by

$$\begin{aligned}\mathcal{B} &= \mathcal{L}_i(s)\mathcal{U}_s(s,\tau)\mathcal{L}_i(\tau)\mathcal{U}_s^{-1}(s,\tau)\mathcal{U}_s^{-1}(t,s)\mathcal{U}_s(t)\rho(0) \\ &= \mathcal{L}_i(s)\mathcal{U}_s(s,\tau)\mathcal{L}_i(\tau)\mathcal{U}_s^{-1}(t,\tau)\mathcal{U}_s(t)\rho(0) \\ &= \mathcal{L}_i(s)\mathcal{U}_s(s,\tau)\mathcal{L}_i(\tau)\mathcal{U}_s^{-1}(\tau)\rho(0) \\ &= \mathcal{L}_i(s)\mathcal{U}_s(s,\tau)\mathcal{L}_i(\tau)\rho_0(-\tau) \\ &= [H_i(s), \mathcal{U}_s(s,\tau)[H_i(\tau),\rho_0(-\tau)].\end{aligned} \tag{S48}$$

Here, $\rho_0(-\tau) \equiv U_s(\tau)\rho(0)$ and the unitary evolution operator $u_s(t)$ is defined by Eq. (S3).

After some mathematical manipulations, $B$ becomes

$$\begin{aligned}\mathcal{B} &= \mathcal{L}_i(s)\mathcal{U}_s(s,\tau)\mathcal{L}_i(\tau)\mathcal{U}_s^{-1}(s,\tau)\mathcal{U}_s^{-1}(t,s)\mathcal{U}_s(t)\rho(0) \\ &= H_i(s)u_s(s,\tau)H_i(\tau)\rho_0(-\tau)u_s^\dagger(s,\tau) - H_i(s)u_s(s,\tau)\rho_0(-\tau)H_i(\tau)u_s^\dagger(s,\tau) \\ &-u_s(s,\tau)H_i(\tau)\rho_0(-\tau)u_s^\dagger(s,\tau)H_i(s) + u_s(s,\tau)\rho_0(-\tau)H_i(\tau)u_s^\dagger(s,\tau)H_i(s),\end{aligned} \tag{S49}$$

and

$$\begin{aligned}&\int_0^t ds\int_0^s d\tau \mathcal{U}_s(t,s)\langle \mathcal{L}_i(s)\mathcal{U}_s(s,\tau)\mathcal{L}_i(\tau)\mathcal{U}^{-1}(s,\tau)\rangle_i \mathcal{U}_s^{-1}(t,s)\mathcal{U}_s(t,0)\rho(0) \\ &= \int_0^t ds\int_0^s d\tau \mathcal{U}_s(t,s)\langle \mathcal{B}\rangle_i \\ &= \int_0^t ds\int_0^s d\tau u_s(t,s)\langle \mathcal{B}\rangle_i u_s^\dagger(t,s).\end{aligned} \tag{S50}$$

From Eqs. (S48)–(S50), Eq. (S47) becomes

$$\begin{aligned}\rho(t) &= [\mathcal{W}^{(2)}]^{-1}(t)\mathcal{U}_s(t,0)\rho(0) \\ &= \mathcal{V}^{(2)}(t)\rho(0) \\ &= \mathcal{U}_s(t)\rho(0) \\ &-\int_0^t ds\int_0^s d\tau u_s(t,s)\langle \mathcal{B}\rangle_i u_s^\dagger(t,s).\end{aligned} \tag{S51}$$

For the two-qubit gate, we use the Hubbard exchange Hamiltonian

$$H_s(t) = \mathcal{J}(t)\boldsymbol{S}_1\cdot\boldsymbol{S}_2 \tag{S52}$$

Then, $u_s(t,\tau) = exp[-i\Phi(t,\tau)\boldsymbol{S}_1\cdot\boldsymbol{S}_2], \Phi(t,\tau) = \int_\tau^t dv\mathcal{J}(v).$ (S53)

## III. Exact Ornstein–Uhlenbeck Validation, Numerical Convergence, and Full-Process Information

In the numerical calculations, we obtain the process maps by averaging exact stochastic Ornstein–Uhlenbeck (OU) trajectories. We use the second-order time-convolutionless (TCL2) approximation only to analyze the two-time structure and to provide a finite-order comparison [1–5,8–10].

### A. Exact stationary OU process

To generate a stationary OU process, we use its exact finite-step transition. The initial value is sampled from the stationary Gaussian distribution. We then generate the subsequent values using independent standard-normal random variables $\eta_n$. We do not use Euler–Maruyama discretization [11,12].

$$\xi_{n+1} = a\xi_n + \sigma\sqrt{(1 - a^2)}\eta_n \tag{S54}$$

$$a = exp(-\Delta t/\tau_c) \tag{S55}$$

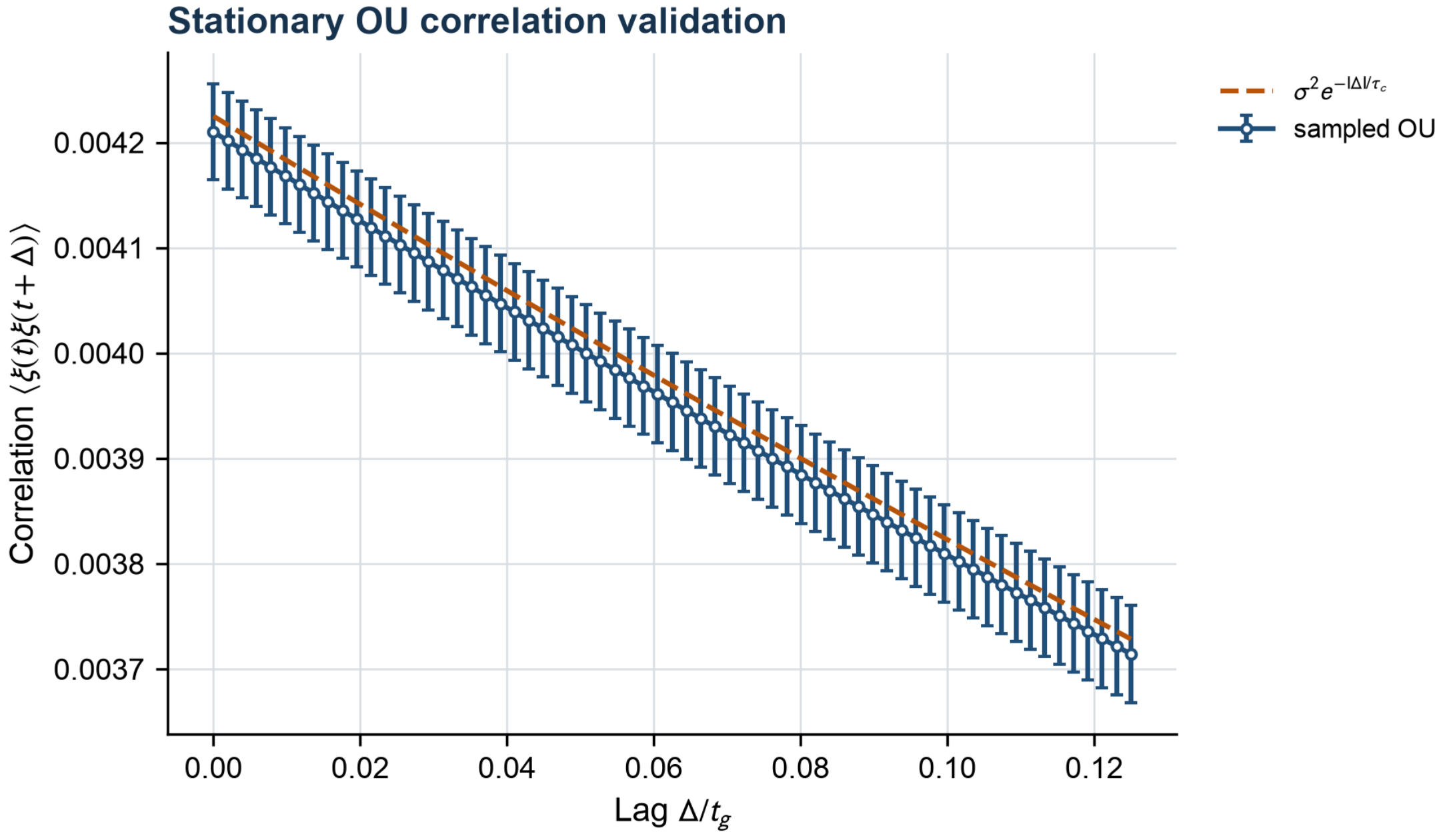

**Supplementary Fig. S1** | Validation of the stationary OU process. Sampled stationary correlation (10,000 independent paths; 512 steps; σ = 0.065 and τc/tg = 1) compared with the exact exponential correlation. Error bars show sampling uncertainty. The sampled variance differs from σ² by approximately 0.35%, and the maximum absolute correlation z-score is approximately 0.35.

### B. Exact trajectory-map construction

For each realization $\xi(t)$, we also use the antithetic realization $-\xi(t)$. By averaging the pair, the leading odd stochastic sampling contributions cancel without changing the OU ensemble. We construct a $4 \times 4$ Pauli-transfer matrix (PTM) for the single-qubit system and a $16 \times 16$ PTM for the two-qubit system [7].

$$\mathcal{E}_{OU} = \langle \mathcal{U}_\xi \rangle = ½\langle \mathcal{U}_\xi + \mathcal{U}_{-\xi} \rangle \tag{S56}$$

Because each exact OU process map is an average of unitary channels, it is completely positive and trace preserving (CPTP) by construction.

The numerical parameters, trajectory counts, time-step settings and random-seed conventions for the single-qubit calculations are summarized in Supplementary Table S1.

## C. Numerical convergence

The numerical parameters, trajectory settings, hidden-case calculations and refined OU/TCL2 comparisons for the two-qubit calculations are summarized in Supplementary Table S2.

To examine numerical convergence, we consider the representative two-qubit point $\lambda = 0.084$ and $r_c = 3.16$. Using 128, 256, 512 and 1024 midpoint steps, we obtain average fidelities of 0.9962621, 0.9961414, 0.9961701 and 0.9962545, respectively. The total variation, approximately $1.2 \times 10^{-4}$, is smaller than the Monte Carlo standard error of approximately $2.3 \times 10^{-4}$. Therefore, at this point, the calculation is limited mainly by stochastic sampling rather than by the propagation time step. We define the average gate fidelity by the standard channel-average relation [13].

The corresponding convergence values, map errors and minimum Choi eigenvalues are given in Supplementary Table S3.

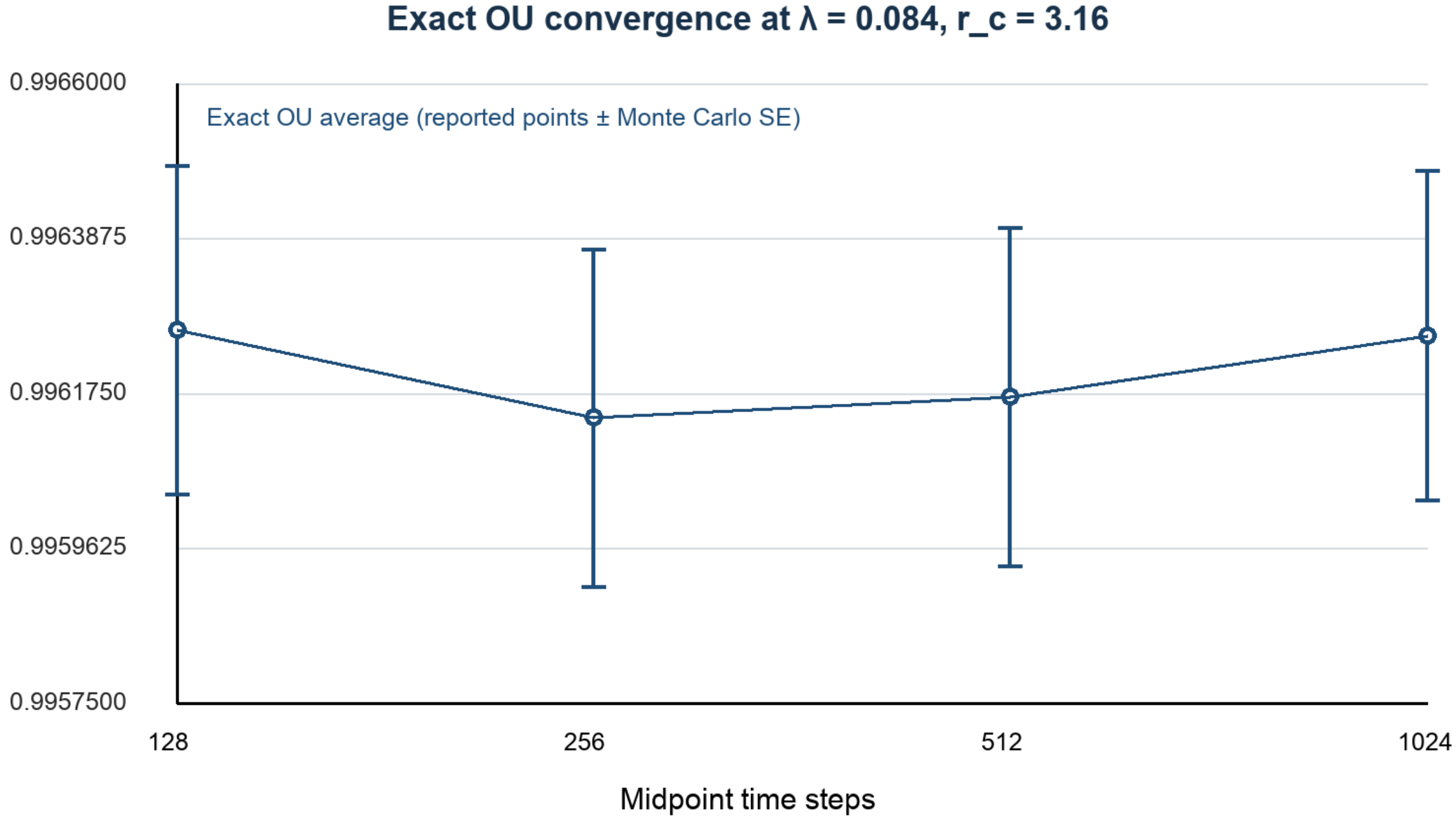

**Supplementary Fig. S2** | Numerical convergence of exact OU process sensitivities. Actual calculated average-fidelity points at λ = 0.084 and rc = 3.16 for 128, 256, 512 and 1024 midpoint steps; whiskers are the reported Monte Carlo standard errors. Lines connect calculated points only as a visual guide.

For the single-qubit derivatives, centered logarithmic steps of 0.03 and 0.06 agree within approximately 0.05%. For 500–2,000 trajectories, the sampling-relative spreads of $D_r$ are approximately 1.5% at $r_c = 0.01$, 4.3% at $r_c = 0.3$, and 7.3% at $r_c = 30$. The relative uncertainty increases in the long-memory region because the correlation-time derivative approaches the Monte Carlo floor.

## D. Exact OU versus TCL2

For t_g = 1, the exact OU and TCL2 results preserve the same qualitative control-history and commuting/noncommuting behavior. At the reference points, they generally agree within approximately 3%. Across the broader tested parameter range, the median relative difference is approximately 3.3%, the 90th percentile is approximately 7.7%, and the maximum reaches approximately 12%. We do not treat TCL2 as exact.

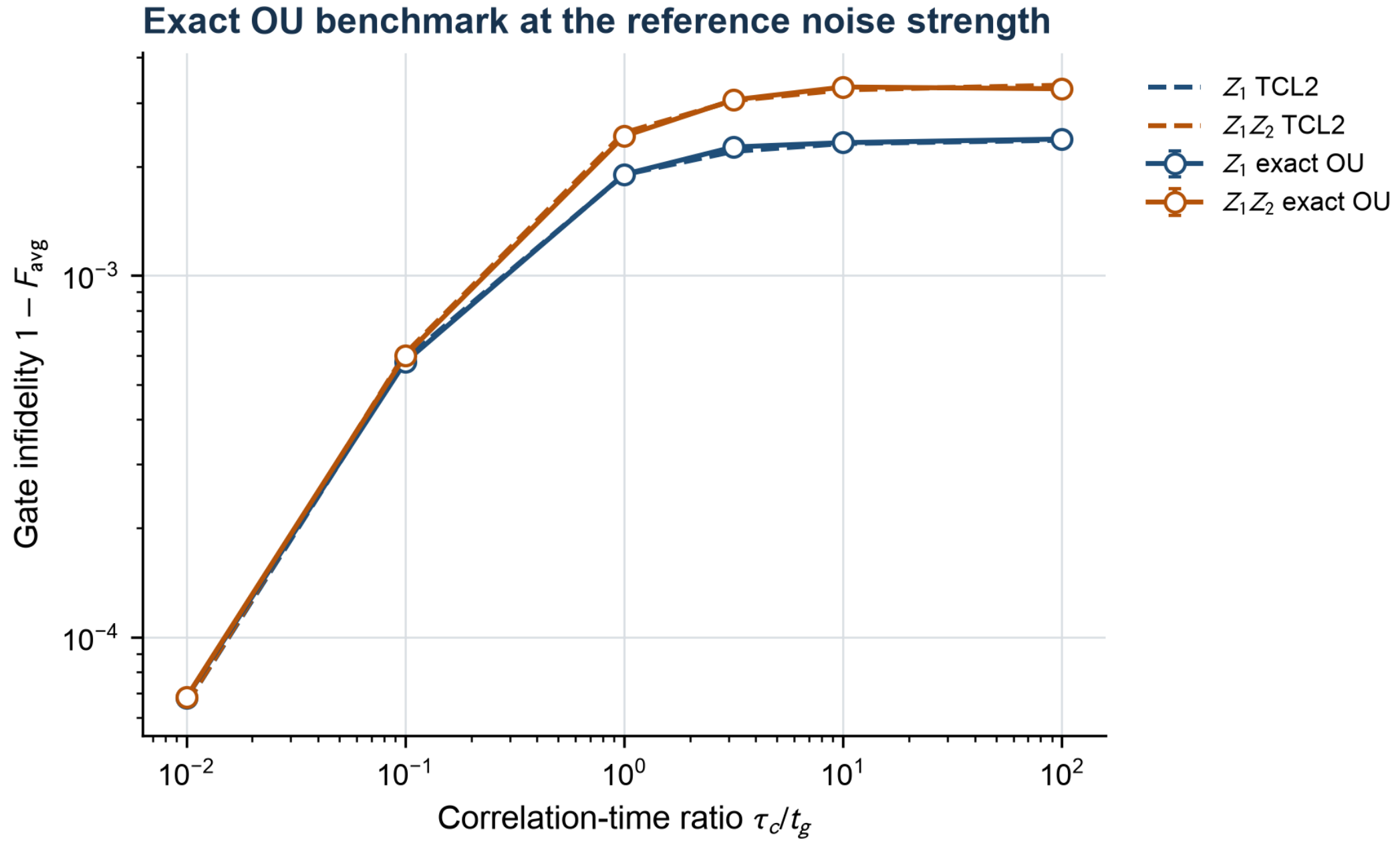

Supplementary Fig. S3 | Exact OU versus TCL2 process validation. Gate infidelity for noncommuting $Z_1$ and commuting $Z_1 Z_2$ stochastic couplings at $\lambda = 0.065$. Markers show exact OU averages and dashed curves show TCL2 predictions; reported Monte Carlo uncertainties are included where visible.

We use TCL2 to understand the analytical two-time structure, while the principal numerical results are obtained from exact stochastic OU trajectories [1–5,8–10].

## E. Complete positivity

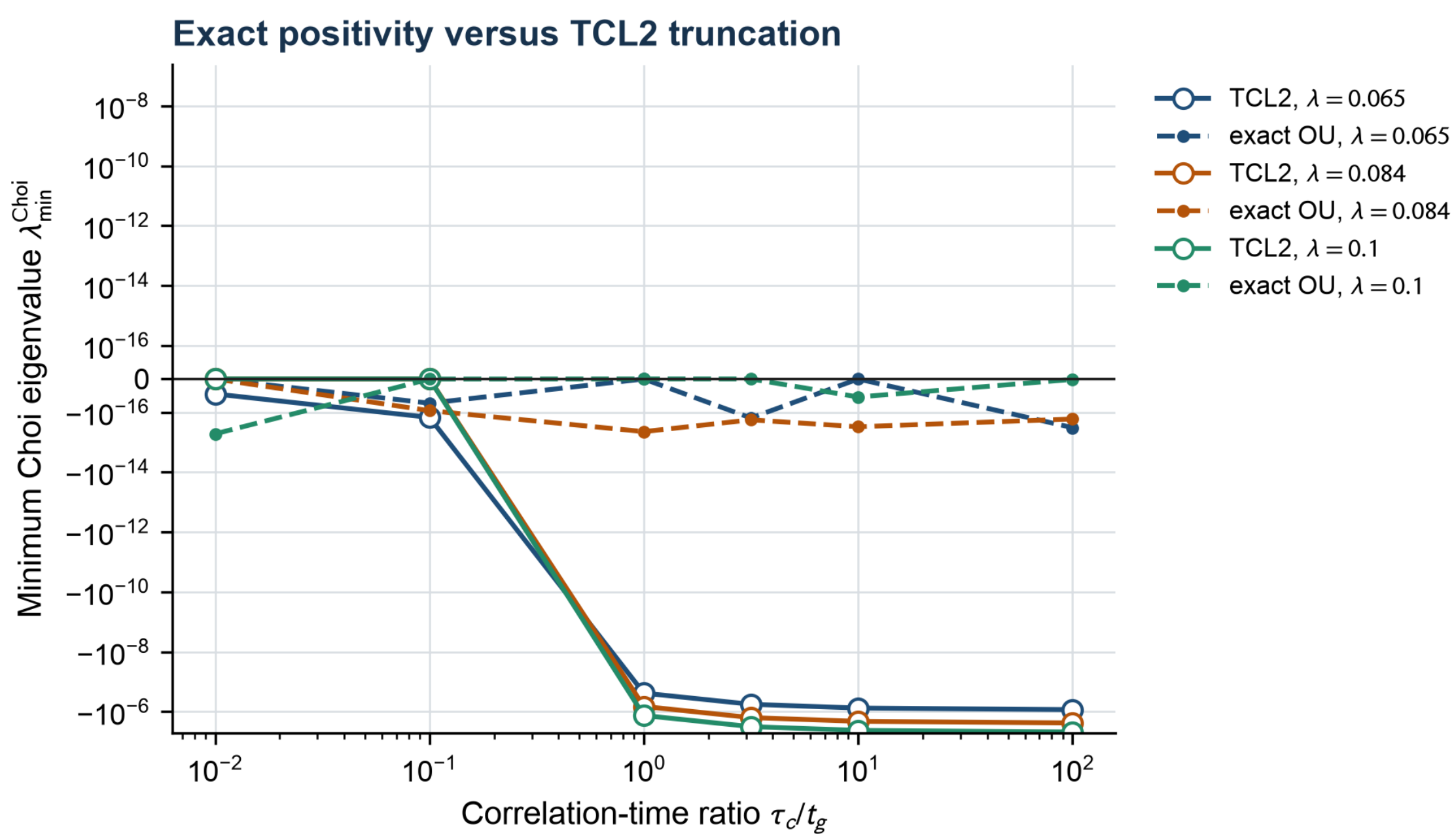

**Supplementary Fig. S4** | Complete positivity of exact OU maps and TCL2 truncation. Minimum Choi eigenvalues for exact OU and TCL2 maps. Exact values remain consistent with floating-point zero (down to approximately $-10^{-15}$), while small

negative TCL2 eigenvalues appear in long-memory, moderate-strength regimes.

Because the exact maps are averages of unitary channels, they are CPTP by construction, and their minimum Choi eigenvalues remain at the level of floating-point zero. In contrast, small negative Choi eigenvalues can appear in the TCL2 maps as finite-order truncation artifacts; they do not indicate a physical loss of complete positivity. We use this comparison only as a consistency check on the finite-order TCL2 approximation [14].

F. Measurement compression relative to the full-process information

We form a genuinely nested observable hierarchy using the same rectangular, smooth-sin-squared, and front-loaded control protocols at every level. Each component uses one scale defined over the complete common parameter grid, and every logarithmic derivative is evaluated with the same grid-consistent rule with respect to log(λ) and log(r_c).

$$R_{ij} = \frac{1}{d} Tr[P_i \, \mathcal{E}(P_j)] \tag{S57}$$

$$J = [\, \partial\, vec\, R / \partial \log \lambda \,,\ \partial\, vec\, R / \partial \log r_c \,] \tag{S58}$$

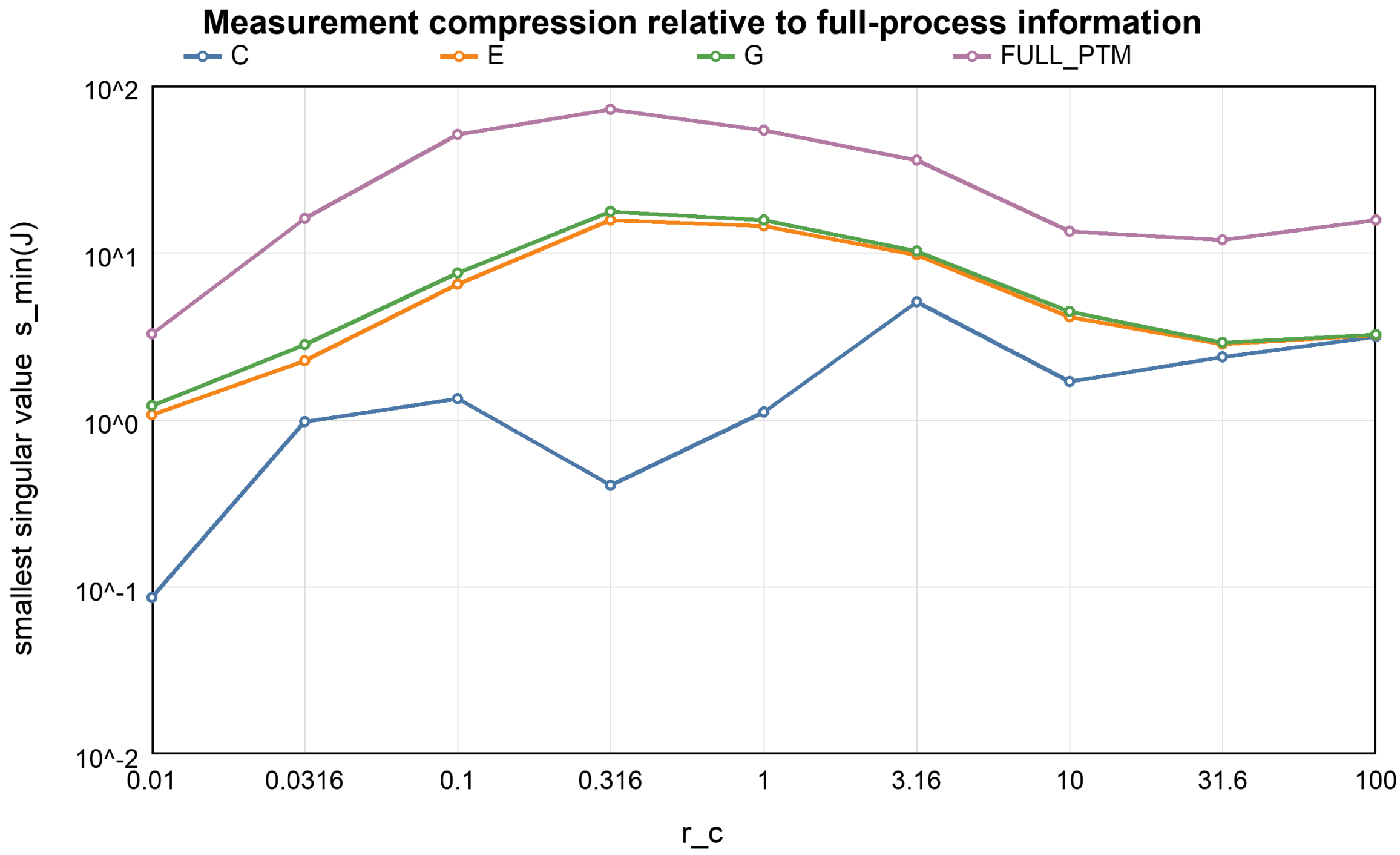

Supplementary Fig. S5 | Measurement compression relative to the full-process information. Smallest singular value of the consistently normalized logarithmic-parameter Jacobian at λ = 0.084 for a nested hierarchy of gate-only observables (C), gate plus one selected PTM coefficient (E), gate plus two selected PTM coefficients (G), and the complete nontrivial PTM representation for the same control protocols (FULL_PTM). All levels use the same derivative and normalization convention.

The consistently normalized nested hierarchy satisfies FULL_PTM ≥ G ≥ E ≥ C at every analyzed point. In the long-memory region (r_c >= 10), the complete informative PTM is itself strongly ill-conditioned, with a median Jacobian condition number of approximately 118.7 over the analyzed points r_c = 10, 31.6, and 100, where the condition number is defined as kappa(J) = s_max(J)/s_min(J). This shows that the loss of independent correlation-time information is not solely a consequence of severe measurement compression; substantial ill-conditioning is already present at the level of the full finite-duration process. This monotonic ordering is consistent with the use of a genuinely nested observable hierarchy under a common normalization and derivative convention.

## IV. Process Observables, Identifiability, and Measurement Compression

In this section, we examine how much of the memory information contained in the full process is retained when only a restricted set of experimentally accessible observables is measured. This analysis provides the measurement interpretation of the main results.

### A. Gate-infidelity degeneracy

A single gate-infidelity measurement does not determine both $\lambda$ and $r_c$ uniquely. Instead, it defines an extended manifold in the $(\lambda, r_c)$ plane. Rectangular, smooth $sin^2$, and front-loaded equal-area exchange histories separate these manifolds because they dress the noncommuting stochastic operator differently. The intersections remain shallow; thus, different controls provide additional information, although gate infidelity alone remains strongly compressed [15,16].

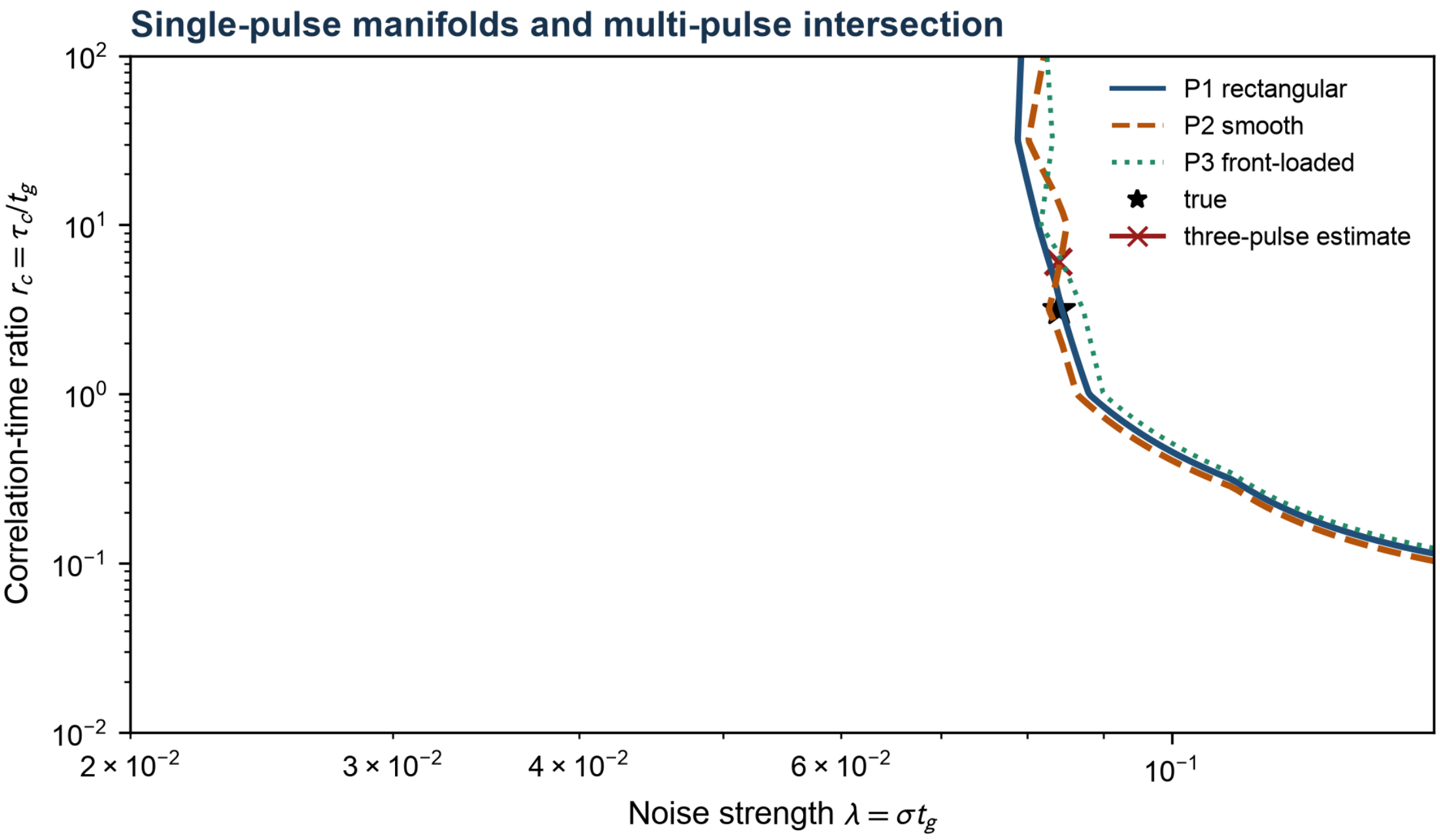

**Supplementary Fig. S6** | Gate-infidelity manifolds and correlation-time degeneracy. Constant-observable manifolds for a representative hidden case (λ = 0.084, rc = 3.16) under rectangular, smooth and front-loaded equal-area pulse histories. The

pulse-dependent manifolds split but intersect at shallow angles.

## B. Confidence-region geometry

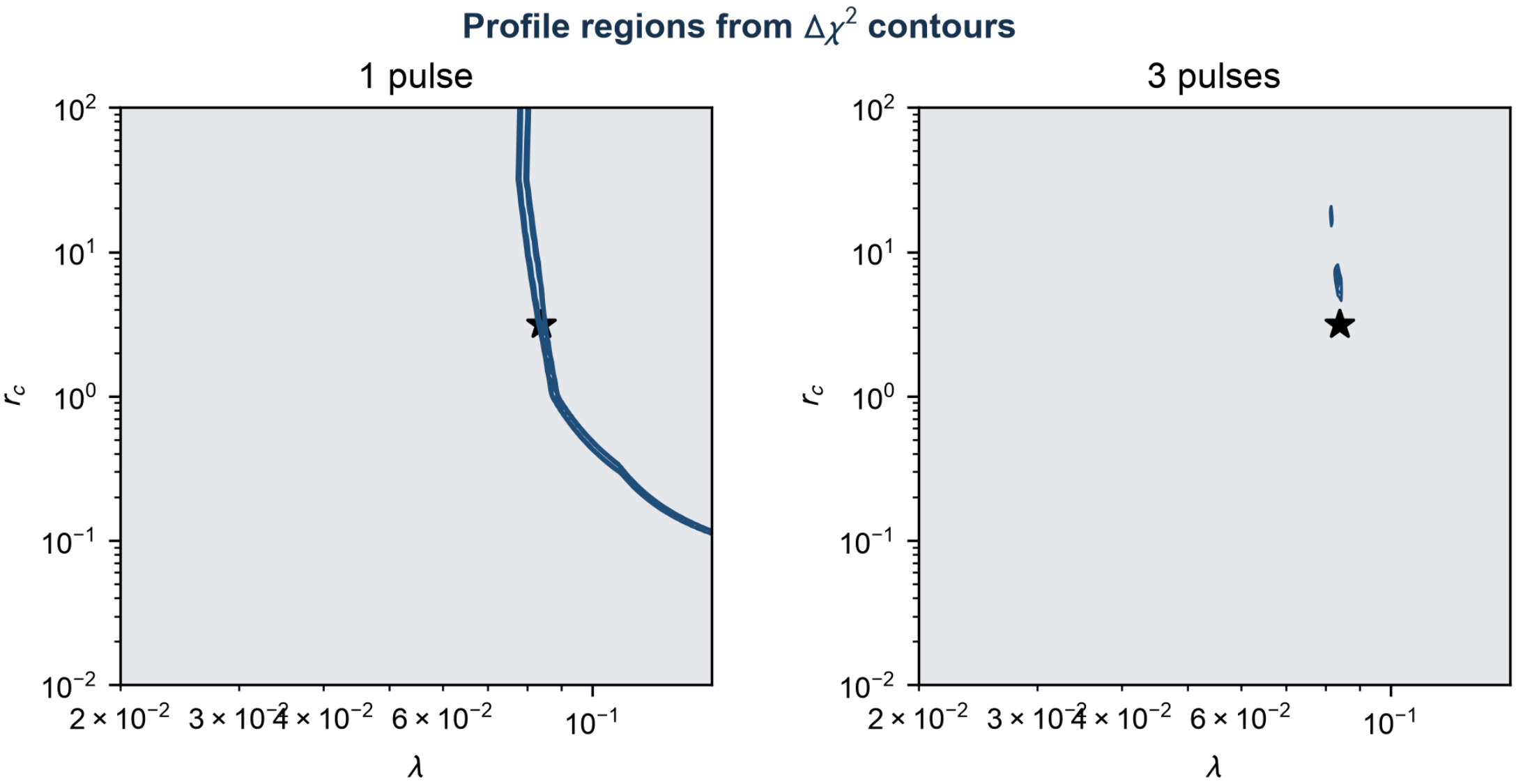

**Supplementary Fig. S7** | Multi-pulse confidence regions for OU parameter reconstruction. Actual $\Delta\chi^2$ profile regions for one pulse and three pulses under the validated 1% relative-error model with a $10^{-5}$ absolute floor. The three-pulse region narrows but remains elongated; no Gaussian ellipse replacement is used.

## C. Reconstruction across hidden cases

We also tested 16 independent off-grid hidden cases. The noiseless success fraction is 0% for one pulse and approximately 6.25% for both two and three pulses. These results show that the noise strength is more readily determined than the correlation time, consistent with the extended and shallow profile geometry. These pulse-count success fractions refer only to gate-infidelity observables; the augmented PTM hierarchy summarized in Supplementary Table S4 is a separate reconstruction analysis.

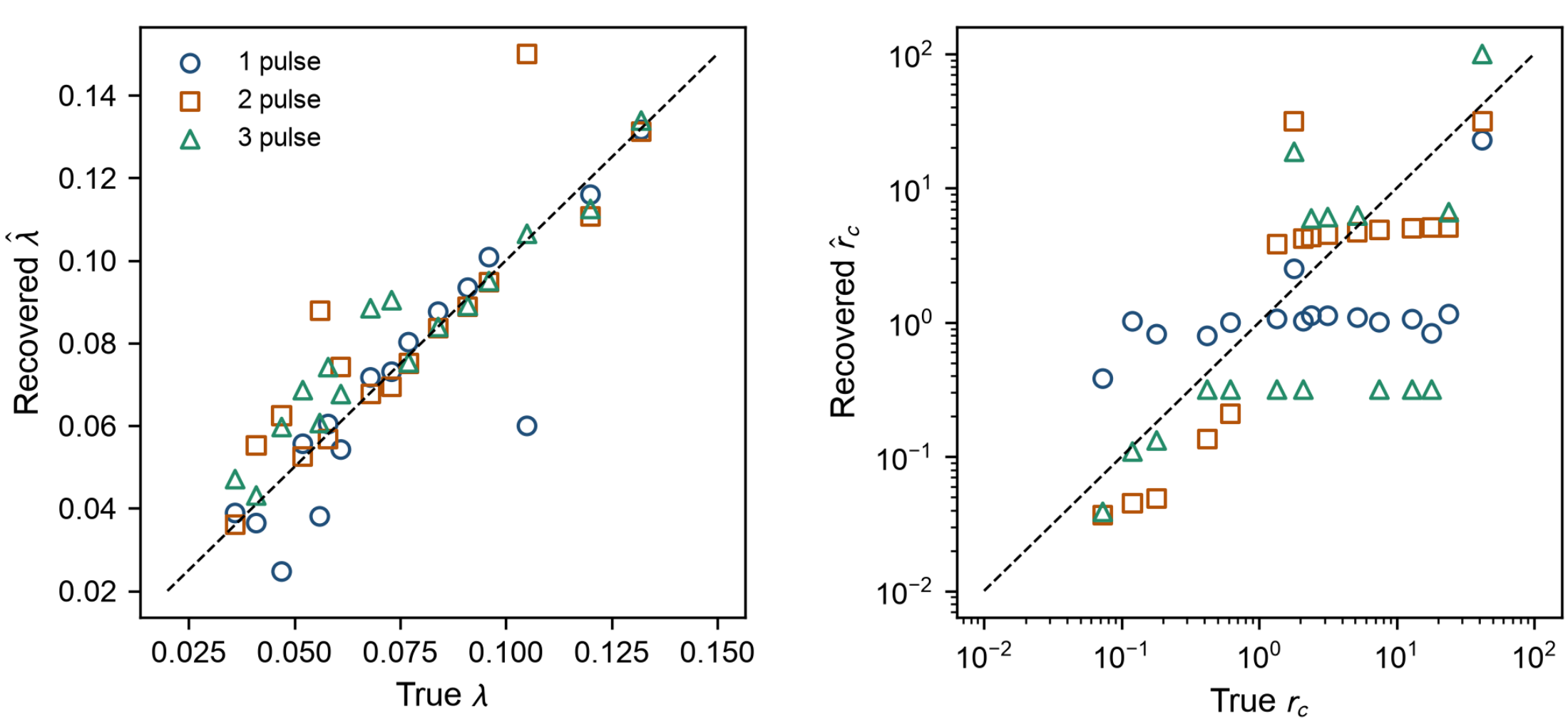

**Supplementary Fig. S8** | Exact hidden-case reconstruction from gate-infidelity observables. Recovered versus true λ and rc for 16 off-grid cases using one, two and three gate-infidelity observables; dashed diagonals denote exact recovery.

## D. Addition of process-resolved PTM information

We first add the process-resolved coefficient $R_{XX,YY}$. This coefficient retains an operator-orientation transfer that is not contained in scalar gate infidelity.

$$R_{XX,YY} = \frac{1}{4}\,\mathrm{Tr}[XX\,\mathcal{E}(YY)] \tag{S59}$$

We then include $R_{XY,YX}$ after conditioning on the gate observables and $R_{XX,YY}$.

$$R_{XY,YX} = \frac{1}{4}\,\mathrm{Tr}[XY\,\mathcal{E}(YX)] \tag{S60}$$

The second coefficient provides only a modest amount of additional information once $R_{XX,YY}$ is included.

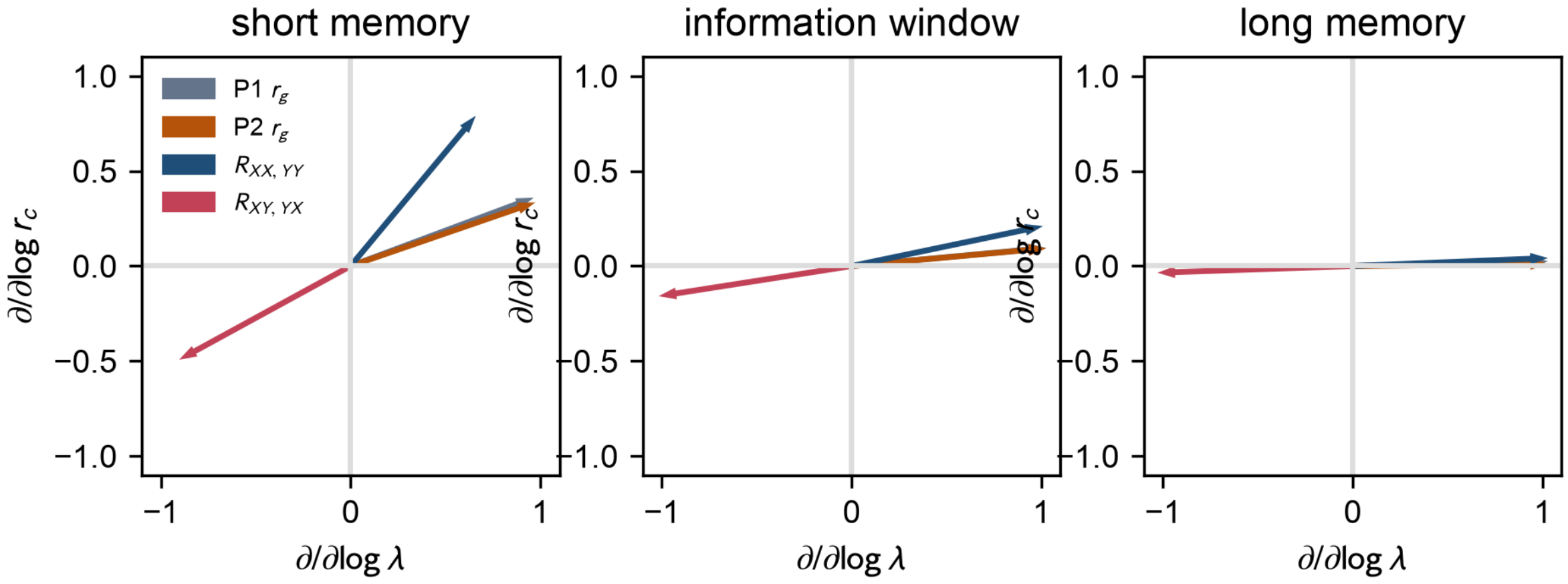

**Supplementary Fig. S9** | Process-resolved sensitivity directions across memory regimes. Normalized logarithmic sensitivity directions for gate and selected PTM observables in short-memory, intermediate-memory and long-memory regimes. Near-parallel long-memory directions expose the residual information ceiling.

## E. Measurement hierarchy

We compare gate-only information, gate plus $R_{XX,YY}$, gate plus $R_{XX,YY}$ and $R_{XY,YX}$, and the full informative PTM. Adding $R_{XX,YY}$ improves identifiability, whereas adding $R_{XY,YX}$ does not give a reliable improvement in the global reconstruction. Among the tested minimal sets, the best result is obtained with two gate histories and $R_{XX,YY}$. The corresponding reconstruction results are summarized in Supplementary Table S4.

## F Experimental accessibility

We obtain a general PTM element from a Pauli input component and a Pauli measurement:

$$R_{ij} = \frac{1}{d}\,\mathrm{Tr}[P_i\,\mathcal{E}(P_j)] \tag{S61}$$

To measure $R_{XY,YX}$, we prepare the four $Y \otimes X$ product-eigenstate sign combinations, apply the channel, and measure $X \otimes Y$ in the local Pauli basis.

The required product-state preparations and local Pauli measurement settings are summarized in Supplementary Table S5.

$$R_{XY,YX} = \frac{1}{4}\,\Sigma s_1, s_2\; s_1 s_2\, \langle X \otimes Y \rangle s_1 Y, s_2 X \tag{S62}$$

## G Commuting null

Finally, we consider isotropic two-qubit exchange with the commuting stochastic interaction $H_i \propto Z_1 Z_2$. Because this interaction commutes with the exchange control, the transformed stochastic operator is time independent.

$$[\,H_s(t), H_i\,] = 0 \;\Rightarrow\; u_s \dagger (t,\tau)\, H_i\, u_s(t,\tau) = H_i \tag{S63}$$

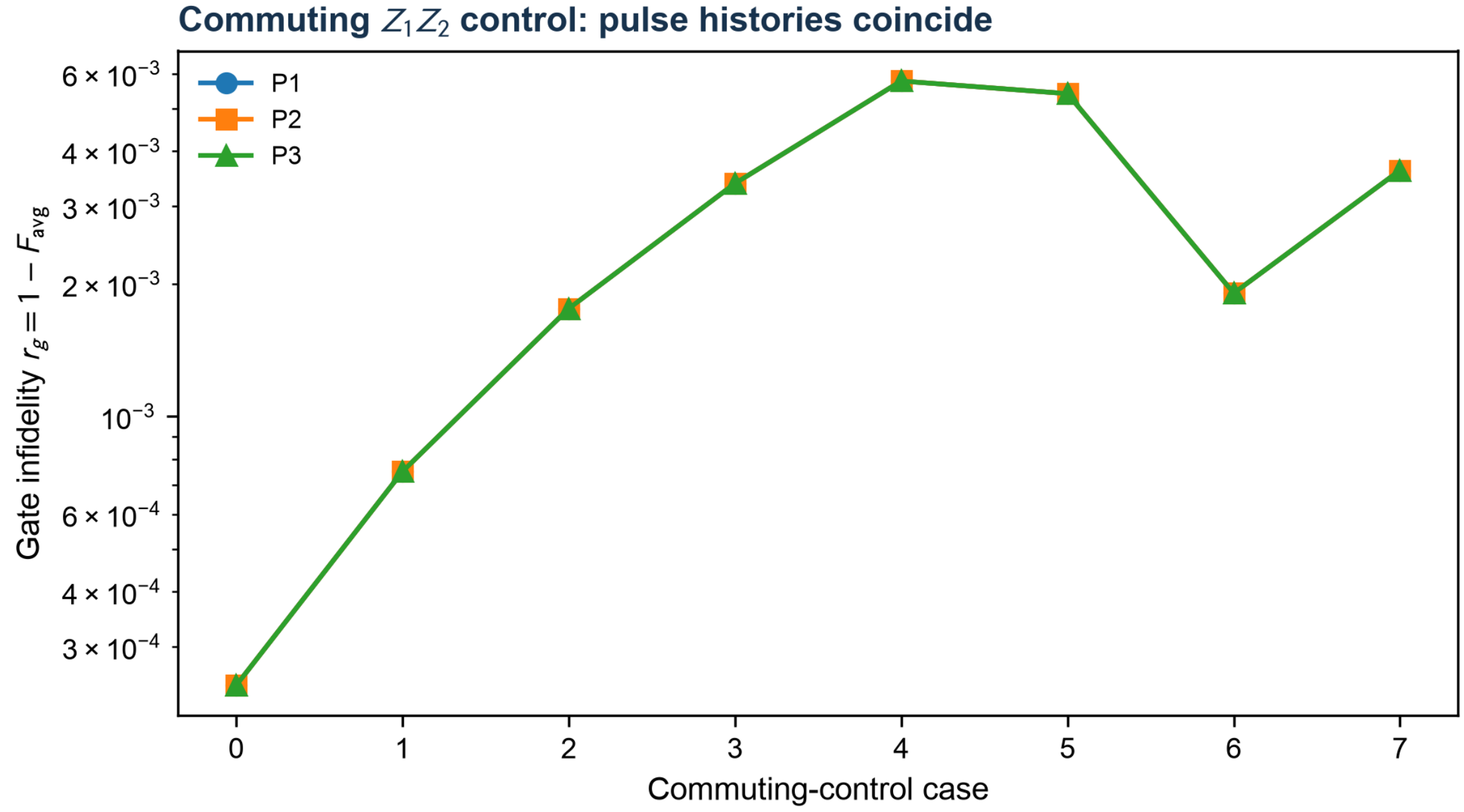

Supplementary Fig. S10 | Collapse of pulse-history information for commuting stochastic coupling. Exact OU gate infidelity for three equal-area pulse histories with $Z_1Z_2$ noise under isotropic exchange. The curves coincide to numerical precision; the validated pulse-to-pulse spread is approximately $10^{-16}$ ($2.22 \times 10^{-16}$ in the multi-pulse gate-infidelity test).

The corresponding PTM null spreads are $1.215 \times 10^{-16}$ for $R_{XX,YY}$ and $1.235 \times 10^{-16}$ for $R_{XY,YX}$. Thus, the pulse histories give the same result to numerical precision. This result shows that the process-information gain originates from the noncommuting transformation of the operator, rather than merely from the use of different pulse envelopes.

**Supplementary Table S1 | Single-qubit OU numerical settings.**

| Quantity | Validated setting |
|---|---|
| $\lambda$ grid | 0.03, 0.05, 0.065, 0.084, 0.10, 0.15 |
| $r_c$ grid | 0.001, 0.003, 0.01, 0.03, 0.1, 0.3, 1, 3, 10, 30, 100, 200, 500 |
| Trajectories | 1,000 (500 antithetic pairs) |
| Time steps | $\max[256, \min(4096, \mathrm{ceil}(4/r_c))]$ |
| Seed family | $20261111 + r_c$ index |

**Supplementary Table S2 | Two-qubit OU parameter grid and exact trajectory settings.**

| Component | Validated setting |
|---|---|
| Forward grid | $7\lambda \times 9r_c \times 3$ pulses; 189 exact points |
| Forward trajectories | 2,000 antithetic trajectories |
| Forward steps | 128 midpoint steps |
| Hidden cases | 16 off-grid cases; 4,000 trajectories; 256 steps |
| Bounds | $0.02 \le \lambda \le 0.15$; $0.01 \le r_c \le 100$ |
| Refined OU/TCL2 points | 10,000 trajectories; 256 steps |
| Seeds | Recorded for every calculation row |

Supplementary Table S3 | Representative convergence values at $\lambda = 0.084$ and $r_c = 3.16$.

| Steps | $F_{\mathrm{avg}}$ exact | $\mathrm{SE}(F_{\mathrm{avg}})$ | Map error | Minimum Choi eigenvalue |
|---|---|---|---|---|
| 128 | 0.9962621 | $2.25\times10^{-4}$ | $5.94\times10^{-4}$ | $-4.05\times10^{-16}$ |
| 256 | 0.9961414 | $2.31\times10^{-4}$ | $1.45\times10^{-3}$ | $-1.17\times10^{-16}$ |
| 512 | 0.9961701 | $2.32\times10^{-4}$ | $1.17\times10^{-3}$ | $-1.58\times10^{-16}$ |
| 1024 | 0.9962545 | $2.26\times10^{-4}$ | $5.60\times10^{-4}$ | $-2.33\times10^{-20}$ |

Note: Map error is the Frobenius-norm difference between the exact OU and TCL2 process maps (map_fro_error) for λ = 0.084, r_c = 3.16, the Z1 rectangular protocol, and the stated step count; values are from ou_exact_convergence.csv (seed 20310821).

**Supplementary Table S4 | Measurement hierarchy and reconstruction performance.**

| Information level | Validated observables | Success | Median $r_c$ error | Interpretation |
|---|---|---|---|---|
| Gate only | $P_1$, $P_2$, $P_3$ gate infidelities | 6.25% | 80.77% | Strong compression |
| Gate + one PTM | $P_1$, $P_2$, $R_{XX,YY}$ | 37.50% | 30.93% | Best tested minimal set |
| Gate + two PTMs | $P_1$, $P_2$, $R_{XX,YY}$, $R_{XY,YX}$ | 37.50% | 31.55% | No robust incremental gain |
| Full informative PTM | Complete nontrivial PTM | — | — | Full-process long-memory ill-conditioning; see Supplementary Fig. S5. |

The reconstruction hierarchy summarized in Supplementary Table S4 is distinct from the nested information hierarchy used in Supplementary Fig. S5. Figure S5 uses a common three-protocol control set at every information level to permit a mathematically nested comparison, whereas Supplementary Table S4 reports the separately validated minimal-observable reconstruction tests described in this section.

Note: Successful reconstruction is defined as simultaneous relative error <10% in λ and <20% in r_c.

**Supplementary Table S5 | Selected PTM observables and minimum experimental settings.**

| Observable | Input preparation | Output measurement | Minimum configurations |
|---|---|---|---|
| $R_{XX,YY}$ | Four $Y \otimes Y$ product-eigenstate combinations | sign $X \otimes X$ local Pauli basis | 4 preparations; 1 basis |
| $R_{XY,YX}$ | Four $Y \otimes X$ product-eigenstate combinations | sign $X \otimes Y$ local Pauli basis | 4 preparations; 1 basis |